\documentclass[fleqn,usenatbib]{mnras}

\usepackage[utf8]{inputenc}

\usepackage{newtxtext,newtxmath}

\usepackage[T1]{fontenc}
\usepackage{ae,aecompl}
\usepackage[mathscr]{euscript}

\usepackage{float}
\usepackage{graphicx}
\usepackage{subfig}
\usepackage{amsmath}
\graphicspath{{./Images/}}
\usepackage{float}
\usepackage{xcolor}

\newcommand{\beq}{\begin{equation}}
\newcommand{\eeq}{\end{equation}}
\newcommand{\beqn}{\begin{eqnarray}}
\newcommand{\eeqn}{\end{eqnarray}}

\defcitealias{Samsing22}{S22}

\usepackage{color}
\definecolor{cerulean}{rgb}{0.0, 0.48, 0.65}

\definecolor{navy}{rgb}{0.2, 0.0, 1.0}

\definecolor{jungle}{rgb}{0.0, 0.5, 0.0}

\makeatletter
\def\@to{to}
\makeatother

\title[Eccentric Mergers in AGN Discs]{Eccentric Mergers in AGN Discs: Influence of the Supermassive Black-Hole on Three-body Interactions}

\author[G. Fabj \& J. Samsing]{
Gaia Fabj,$^{1}$\thanks{E-mail:  gaia.fabj@nbi.ku.dk}
Johan Samsing$^{1}$
\newauthor
\\
$^{1}$Niels Bohr International Academy, The Niels Bohr Institute,
Blegdamsvej 17, 2100 Copenhagen, Denmark  
}

\date{Accepted XXX. Received YYY; in original form ZZZ}

\pubyear{2024}

\begin{document}
\label{firstpage}
\pagerange{\pageref{firstpage}--\pageref{lastpage}}
\maketitle

\begin{abstract}
There are indications that stellar-origin black holes (BHs) are efficiently paired up in binary black holes (BBHs) in Active Galactic Nuclei (AGN) disc environments, which can undergo interactions with single BHs in the disc. Such binary-single interactions can potentially lead to an exceptionally high fraction of gravitational-wave mergers with measurable eccentricity in LIGO/Virgo/KAGRA. We here take the next important step in this line of studies by performing post-Newtonian N-body simulations between migrating BBHs and single BHs set in an AGN disc-like configuration, with a consistent inclusion of the central supermassive black hole (SMBH) in the equations of motion. With this setup, we study how the fraction of eccentric mergers varies in terms of the initial size of the BBH semi-major axis relative to the Hill sphere, as well as how it depends on the angle between the BBH and the incoming single BH. We find that the fraction of eccentric mergers is still relatively large, even when the interactions are notably influenced by the gravitational field of the nearby SMBH. However, the fraction as a function of the BBH semi-major axis does not follow a smooth functional shape, but instead shows strongly varying features that originate from the underlying phase-space structure. The phase-space further reveals that many of the eccentric mergers are formed through prompt scatterings. Finally, we present the first analytical solution to how the presence of an SMBH in terms of its Hill sphere affects the probability for forming eccentric BBH mergers through chaotic three-body interactions.

\end{abstract}


\section{Introduction} \label{sec:Introduction}

The observations of gravitational-wave (GW) sources with LIGO/Virgo/KAGRA (LVK) are ongoing and have to date revealed sources of both
binary black holes (BBHs) \citep{2016PhRvL.116f1102A, 2016PhRvL.116x1103A, 2016PhRvX...6d1015A,
2017PhRvL.118v1101A, 2017PhRvL.119n1101A,  2019arXiv190210331Z, 2019arXiv190407214V}, binary neutron stars (BNSs) \citep{2017PhRvL.119p1101A},
as well as possible black-hole neutrons star mergers \citep{2023PhRvX..13d1039A,2021MNRAS.506.5345H,Vyntantheya22}. However, how and where these sources form in our Universe remain major
unsolved problems. Many formation channels have been proposed, including
field binaries \citep{2012ApJ...759...52D, 2013ApJ...779...72D, 2015ApJ...806..263D, Kinugawa2014MNRAS,2016ApJ...819..108B,
2016Natur.534..512B, 2017ApJ...836...39S, 2017ApJ...845..173M, 2018ApJ...863....7R, 2018ApJ...862L...3S},
dense stellar clusters \citep{2000ApJ...528L..17P, Lee:2010in,
2010MNRAS.402..371B, 2013MNRAS.435.1358T, 2014MNRAS.440.2714B,
2015PhRvL.115e1101R, 2016PhRvD..93h4029R, 2016ApJ...824L...8R,
2016ApJ...824L...8R, 2017MNRAS.464L..36A, 2017MNRAS.469.4665P, Samsing18, 2018MNRAS.481.4775D, 2018MNRAS.tmp.2223S, 2019arXiv190711231S,2022ApJ...931..149R},
active galactic nuclei (AGN) accretion discs \citep{2017ApJ...835..165B,  2017MNRAS.464..946S, 2017arXiv170207818M, 2019arXiv191208218T,2023MNRAS.521..866R},
galactic nuclei (GN) \citep{2009MNRAS.395.2127O, 2015MNRAS.448..754H,
2016ApJ...828...77V, 2016ApJ...831..187A, 2016MNRAS.460.3494S, 2017arXiv170609896H, 2018ApJ...865....2H},
very massive stellar mergers \citep{Loeb:2016, Woosley:2016, Janiuk+2017, DOrazioLoeb:2017},
and single-single GW captures of primordial black holes \citep{2016PhRvL.116t1301B, 2016PhRvD..94h4013C,
2016PhRvL.117f1101S, 2016PhRvD..94h3504C}. To disentangle these different scenarios using GWs, we have to understand how the
observable parameters differ between the different channels. For example, studies indicate that one can distinguish 
dynamically induced mergers from isolated binaries, by analyzing the relative spin orientation of the
merging black-holes (BHs) \citep{2016ApJ...832L...2R}, the orbital eccentricity at some
reference GW frequency \citep{2006ApJ...640..156G, Samsing14, 2017ApJ...840L..14S, Samsing18a, Samsing2018, Samsing18, 2018ApJ...855..124S,
2018MNRAS.tmp.2223S, 2018PhRvD..98l3005R, 2019ApJ...871...91Z, 2019PhRvD.100d3010S, 2019arXiv190711231S},
as well as the mass spectrum \citep{2017ApJ...846...82Z}.
The environment in which the BBH was formed
and merged can also be imprinted in the GW form, showing up as e.g. a GW phase-shift \citep[e.g.][]{2017PhRvD..96f3014I, 2020PhRvD.101h3031D,2024arXiv240305625S,2024arXiv240804603H}. Other probes of formation include e.g.
stellar tidal disruptions \citep[e.g.][]{2019PhRvD.100d3009S, 2019ApJ...877...56L, 2019ApJ...881...75K}.
From these studies it generally follows that dynamically formed mergers tend to have mass ratios near one \citep[e.g.][]{2018PhRvD..98l3005R}, random
spin orientations \citep[e.g.][]{2016ApJ...832L...2R}, as well as a non-negligible fraction of sources with measurable eccentricity in LISA \citep{2018MNRAS.tmp.2223S,2018MNRAS.481.4775D, 2019PhRvD..99f3003K},
DECIGO/Tian-Qin \citep[e.g.][]{2017ApJ...842L...2C, 2019arXiv190711231S}, and LVK \citep{Samsing18}.
This is in contrast to isolated binary mergers, which are more likely to have a non-random spin distribution \citep[e.g.][]{2000ApJ...541..319K, Zaldarriaga2018MNRAS,Hotokezaka2017ApJ,Piran2020ApJ},
larger mass ratios, and eccentricity $\approx 0$ near LISA, DECIGO/Tian-Qin, and LVK.

However, some of the observed GW mergers start to challenge these classical formation channels. For example, GW190521 \citep{GW19a, GW19b} seems to have masses that
are either above the so called BH mass gap \citep[e.g.][]{2020ApJ...904L..26F}, or one above and one below \citep{2020ApJ...891L..31F,2021ApJ...907L...9N}, which certainly was not generally expected from any models prior to this observation. In addition, GW190521 suggests better consistency with an eccentric waveform with $e\ge 0.1$ at 10 Hz, rather than a quasi-circular one \citep[e.g.][]{2020ApJ...903L...5R,2022ApJ...940..171R,2022NatAs...6..344G}, and a possible corresponding electro-magnetic (EM) counterpart \citep[e.g.][]{Graham20}. These
features have led to the proposal that GW190521 could have formed in an AGN disc, where BHs can grow to high masses
through gas accretion \citep[e.g.][]{2012MNRAS.425..460M,2022ApJ...928..191G} or repeated mergers \citep{Tagawa21}. They can also encounter each other in disc-like configurations that have been shown to produce
up to two orders of magnitude more eccentric mergers compared to isotropic environments \citep{Samsing22} (from hereon referred to as \citetalias{Samsing22}), as well as interact with the surrounding
gas to create possible EM counterparts \citep[e.g.][]{Graham23}. 

This observation, the underlying rich physics that brings together stellar-origin black holes,
supermassive black holes, gas, dynamics and disc-accretion physics, have opened up a wealth of studies on how BBHs might form and merge in such
AGN-disc environments \citep{Tagawa20}. 
The general picture for this formation channel is that BHs can be formed through different mechanisms. They can be captured by the disc \citep[e.g.][]{2017ApJ...835..165B,Fabj20, Macleod20, Nasim22, Generozov22, 2024MNRAS.528.4958W}, form in the disc as a result of star formation in the outskirts \citep{2017MNRAS.464..946S,2019MNRAS.487.1675A}, or they can be the result of core collapse of stars that undergo rapid mass accretion in high-density regions of the accretion disc  \citep{Jermyn21,Cantiello21,2021ApJ...916...48D,2022ApJ...929..133J,2023ApJ...946...56D}. 

We envision this population of black-holes to then migrate through the disc during which they may pair up
to form binaries. These binaries are either brought to merge in the disc through a combination of gas drag and GW radiation 
\citep{Tagawa20,Li22a,2022ApJ...940..155D, 2023MNRAS.522.1881L, Li22c,2024ApJ...964...61D,2024ApJ...970..107C}, or surviving inside the
disc for them to later encounter either other BBHs or single BHs \citep{Tagawa20, Samsing22}. 
In \citetalias{Samsing22} it was argued that 3-body interactions, i.e. interactions between assembled
BBHs and single BHs in the disc dominate the formation of BBH mergers, as each interaction hardens the BBH, possibly increase its eccentricity, and thereby bring the BBHs to merge on a short timescale.
In \citetalias{Samsing22}, the first AGN-disc like 3-body simulations were performed with the inclusion of Post-Newtonian (PN) radiation terms \citep[e.g.][]{Blanchet06, Blanchet14}, which showed that such interactions
lead to an exceptionally high fraction of eccentric mergers. 

Other studies focusing on the formation process of the BBH itself under the influence of an SMBH, through
processes known as Jacobi captures, have also found that these BHs tend to pair up and possibly merge with very high eccentricity \citep{Boekholt23}.
Recent work \citep{Trani23} has performed complementary studies on how BHs interact in stellar-remnant discs environments that do not necessarily move on Keplerian
orbits as expected in AGN-discs, and find that eccentric mergers are formed in large fraction.  
Recent studies have started to explore the effect of a gaseous disc in the problem of BH pairing
with the inclusion of gas \citep{ 2023ApJ...944L..42L, Delaurentiis23,2024MNRAS.533.1766W,2023MNRAS.521..866R,Rowan23b,Rowan24a}. 
While these studies are
encouraging, the problem of gas friction exerted by the accretion disc when modeling the PN 3-body scattering problem in AGNs is a rather challenging task. 
Therefore, in this work we neglect the role of gas and focus on the other prominent  environmental effect in AGNs: the tidal field exerted by the SMBH.

In this paper we take the next step in building up our understanding of how BBHs might be brought to merger in disc-like environments, by performing
PN simulations of single BHs interacting with BBHs in disc-like configurations
with the inclusion of the central massive black hole in the equation-of-motion.
This is a necessary extension to the work by \citetalias{Samsing22},
who did not include the effect from the tides due to the SMBH in the scatterings. In addition, we adopt a set of initial conditions (ICs) that are closer to the ones
that are expected in disc-like environments, where the single and binary slowly approach each other until they interact through their common Hill sphere.
The question that naturally arises is whether the tidal field and the more constrained ICs drastically decrease the high number of eccentric mergers found in the aforementioned study,
or whether instead eccentric mergers are a robust indicator of BBHs assembled in disc-like environments.
In this paper we address the previous questions, present results that are directly comparable to the ones from \citetalias{Samsing22}, and introduce new
important elements to acquire a better understanding of this problem.

With these motivations, we start in $\S$\ref{sec:Formation of Eccentric Mergers} by reviewing the theory of assembling eccentric mergers through binary-single interactions taking place
in disc-like environments. We then in $\S$ \ref{sec:tid fields} extend this theory to include the effect from a tidal boundary caused by the presence
of the nearby SMBH. After this, we move onto performing PN interactions between a BBH and single BH all orbiting a SMBH ($\S$ \ref{sec:single}), for which we present outcome distributions ($\S$ \ref{sec:class} \& $\S$ \ref{sec:orbres}) and phase-space diagrams in $\S$ \ref{sec:phase}. We especially focus on the outcome of probabilities for eccentric mergers as a function of
BBH semi-major axis (SMA) relative to the size of the Hill sphere, as well as how the results depend on the inclination angle between the BBH and the incoming single BH ($\S$ \ref{sec:3D}). 
Finally, we conclude and highlight future directions in this problem.

\section{Three-body Interactions in a Tidal Field} \label{sec:Methods}

\begin{figure*}
\centering
\includegraphics[width=\linewidth]{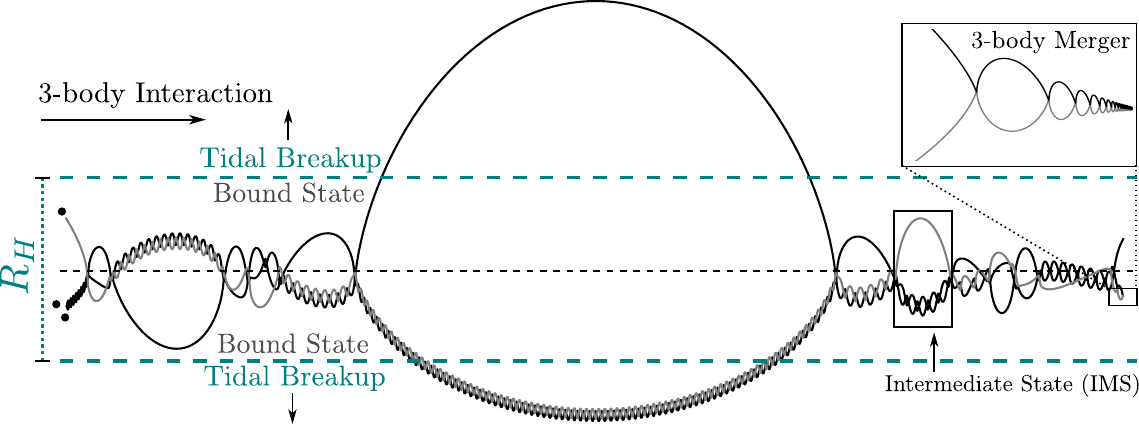}
\caption{Illustration of binary-single interaction propagating from left to right resulting in a merger between two of the three objects while the remaining single is still bound (see Sec. \ref{sec:Formation of Eccentric Mergers}). The system goes through several intermediate states (IMS) before GW capture of two of the three objects. 
For illustrative purposes, we include horizontal dashed lines to indicate the size of a Hill sphere ($R_{\rm H}$) in case the interaction takes place in the presence of a tidal field. If the interaction extends beyond $R_H$, the system becomes unbound and undergoes tidal breakup. The tidal boundary therefore limits the number of interactions $\mathcal{N}$ and it  increases the probability of breaking up the system before natural completion. This consequently leads to a reduction in the probability for 3-body and eccentric mergers. The derived correction factor is shown in Eq. \ref{eq:correction}.
\label{fig:BSscattering}}
\end{figure*}

The three-body scattering problem has been studied in great detail with different physical motivations, from the context of classical cluster evolution \citep{Heggie:1975uy}
to including PN correction terms to follow the formation of GW driven mergers of compact objects \citep{Samsing14, Samsing18}.
While several theoretical studies have been put forward for describing especially the fraction of eccentric merges forming
in chaotic 3-body interactions \citep[e.g.][]{Samsing14, 2016PhRvD..93h4029R, Samsing18}, only a few have extended this theory to include how a nearby perturber, such as an SMBH, affects the range, fraction, and nature of
outcomes \citep{2018MNRAS.474.5672L,2019ApJ...885..135T,Ginat21,Trani23,2023arXiv231003801R}. 
In \citetalias{Samsing22} it was shown analytically and numerically that co-planar scatterings, such as those that might take place in AGN-disc environments,
naturally give rise to an exceptionally high fraction of eccentric GW mergers due to the controlled geometric setup, however neglecting the effect of the SMBH.
Below we start by reviewing the relevant 3-body scattering theory,
after which we derive how the presence of a nearby object to leading order has an effect on the fraction of eccentric mergers forming during the chaotic state of the 3-body system.
This not only adds an important component to this theoretical framework, but also sets the stage and provides motivation for PN scatterings with an SMBH that we present later
in the paper.

\subsection{Formation of eccentric mergers}\label{sec:Formation of Eccentric Mergers}

We start by considering Fig. \ref{fig:BSscattering}, which shows the evolution of an isolated binary-single scattering, i.e. without a nearby perturber, that goes through several resonances, or intermediate
binary-single states (IMS), before one of them undergoes a GW inspiral merger while being bound to the remaining single.
This outcome we refer to as a 3-body merger \citep{Samsing18}. The condition for this to happen is that the inspiral time of the IMS binary, $t_{b}$, has to be less than the time for the remaining single
to return, $T_{bs}$. In the high eccentricity limit the merger time of the IMS binary can be approximated by \citep{Peters64},
\begin{align}
	t_m \approx \left( \frac{2^{7/2}5c^5}{512G^3} \right) a_b^{1/2} m^{-3} r_m^{7/2}
    \label{eq:t_m}
\end{align}
where $a_b$, $m$ and $r_m$, denote the initial IMS binary semi-major axis, 
individual black hole mass (throughout the paper we assume the triple system to be equal mass), and pericenter distance, respectively. 
The timescale of the binary-single orbital time $T_{bs}$ scales with the orbital time of the initial binary such that,
\begin{equation}
T_{bs} \sim 2\pi\sqrt{\frac{a_b^3}{3Gm}},
\end{equation}
up to a constant factor that does not play a role here for our purpose. By setting $t_m$ equal to $T_{bs}$, one can isolate for the critical
pericenter distance, $r_m$, that the two IMS binary objects need to have upon formation for their merger time to be short enough
for them to merge while still being bound to the single. This leads to:
\begin{align}
	r_m   & \approx \left(\frac{512 \pi \sqrt{3}G^{5/2}}{2^{5/2}5c^{5}}\right)^{2/7} a_b^{2/7} m^{5/7} \approx \mathscr{R} \times (a_{b}/ \mathscr{R})^{2/7},
\label{eq:r_m}
\end{align}
where $\mathscr{R}$ here denotes the Schwarzschild Radius of one of the BHs with mass $m$. To get a sense of the scales, one
finds that $r_m/\mathscr{R} \approx 100$, for $m = 20M_{\odot}$ and $a_b = 1\ \rm AU$, i.e. if the IMS binary is formed with a pericenter distance  $<100 \times \mathscr{R}$
then it is likely to undergo a GW capture merger for this setup \citep{Samsing18}.

To estimate the actual probability that a given IMS binary undergoes such a merger, we have to consider the distribution of
SMA and eccentricity for a given IMS, as $r_m$ is directly linked to these quantities through $r_m = a_b(1-e_b)$.
In the equal mass case, the binary SMA, $a_{b}$, is of order the initial SMA, $a_0$, and we (for this part) therefore assume that
\begin{equation}
a_b \approx a_0.
\end{equation}
The IMS binary eccentricity, $e_b$, on the other hand can vary greatly \citep{1976MNRAS.177..583M, 2006tbp..book.....V, 2019Natur.576..406S} and in the co-planar chaotic limit follows the distribution,
\begin{equation}
P(e) \approx e/\sqrt{1-e^2}.
\end{equation}
This distribution implies that the probability that an IMS binary in the co-planar limit forms with a pericenter distance $< r$ is given by,
\begin{equation}
P( < r) \approx \sqrt{2r/a_0},
\label{eq:Pltr}
\end{equation}
which in the case of an IMS binary merger set by the condition $r < r_m$ translates to the relation,
\begin{equation}
P(r < r_m) \approx \sqrt{2r_m/a_0} \approx \left(a_0/\mathscr{R}\right)^{-5/14}
\end{equation}
where we have used the notation from Eq. \ref{eq:r_m}. Now, this is the probability that a given IMS undergoes a merger in the co-planar limit.
To get the full probability that the entire binary-single interaction results in a 3-body merger,
we have to take into account that the interaction goes through several IMS, a number we denote $\mathcal{N}$ \citep{Samsing14}.
In the limit where the probability for a single IMS to undergo a merger is $\ll 1$, the total probability for a 3-body merger is simply the probability for
one IMS binary to undergo a merger, i.e. $P(r < r_m)$, times the number of intermediate states, $\mathcal{N}$,
\begin{align}
	P_{3b} & \approx \mathcal{N} \times \left(\frac{512 \pi \sqrt{3}G^{5/2}}{2^{-1}5c^{5}}\right)^{1/7} a_0^{-5/14} m^{5/14}\nonumber\\
		   &  \approx \mathcal{N} \times \left(a_0/\mathscr{R}\right)^{-5/14},
	\label{eq:P3b}
\end{align}
where $\mathcal{N} \approx 20$ as discussed further in the sections below.

For a 3-body merger that is always highly eccentric at formation, the corresponding GW frequency where most of the power is emitted, also referred as the \textit{GW peak frequency} \citep{DJD18}, is related to the binary pericenter $r_f$ distance as,
\begin{equation}
f_{\rm gw} \approx \pi^{-1}\sqrt{2Gm/r_f^{3}},
\end{equation}
which implies that if an IMS binary forms with a pericenter less than
\begin{equation}
r_f \approx \left(\frac{2Gm}{{\pi}^2 f_{\rm gw}^2}\right)^{1/3},
\label{eq:rf}
\end{equation}
it consequently shows up eccentric at GW frequency that is higher than $f_{\rm gw}$. 
It might be that the merger forms with a GW peak frequency that is below the observable band of LVK \citep{Samsing14}.
In this case, one has to evolve the binary until it reaches the band through the emission of GWs; however, at this point the eccentricity might be too low to be measured. Therefore,
3-body mergers are not necessarily representative of the fraction of observable highly eccentric mergers in e.g. LVK as discussed in \citet{Samsing18}.
For an IMS merger to clearly show up as an eccentric merger, its GW peak frequency at formation has to be near the observable band.

With the critical distance from Eq.~\ref{eq:rf}, one can easily estimate the probability for the outcome in a similar way to what we did in Eq. \ref{eq:P3b}, which is $\approx \mathcal{N} \times \sqrt{2r_f/a_0}$.
In the more general case for the IMS binary to appear at GW frequency $f_{\rm gw}$ with an eccentricity $e_f$, the probability should instead be based on the critical pericenter distance,
\begin{equation}
r_{e,f} \approx \left( \frac{2Gm}{f_{\rm GW}^2 {\pi}^2}\right)^{1/3} \frac{1}{2}  \frac{1+e_{f}}{e_{f}^{12/19}} \left[ \frac{425}{304} \left(1 + \frac{121}{304}e_{f}^2 \right)^{-1} \right]^{870/2299}.
\end{equation}
as discussed in \citetalias{Samsing22}. Here the first term is simply $r_f$ from Eq. \ref{eq:rf}, which implies that the ratio $r_{e,f}/r_f$ only depends on the eccentricity threshold at $f_{\rm gw}$. For example, for $e_f = 0.1$, $r_{e,f}/r_f \approx 3$, meaning that if the IMS binary is assembled with a pericenter distance that is $\approx 3$ times larger than
$r_f$, it appears with an eccentricity $0.1$ at GW peak frequency $f_{\rm gw}$.
Generally, one should note that for an eccentric source to also merge the pericenter has to be less than $r_{m}$, which indeed is the case for
most astrophysical systems near the LVK band.

With relevant astrophysical scales we can summarize these analytical estimations as
\begin{equation}
P_{3b} \approx 0.15 \left[\frac{m}{20M_{\odot}} \right]^{5/14} \left[\frac{a_0}{1\ {\rm AU}} \right]^{-5/14},
\end{equation}
and
\begin{equation}
P_{3b}(e > 0.1: > 10\ {\rm Hz}) \approx 0.15 \left[\frac{m}{20M_{\odot}} \right]^{1/6} \left[\frac{a_0}{1\ {\rm AU}} \right]^{-1/2},
\end{equation}
as further explained in \citetalias{Samsing22}.
The question is how these general results, which also have shown excellent agreement with numerical simulations \citepalias{Samsing22}, change when
the chaotic three-body interactions take place in a tidal field from a nearby perturber, such as an SMBH. Below we estimate how this addition affects the outcomes.

\subsection{Effects from tidal fields}\label{sec:tid fields}

The effect from a nearby object shows up (radially) to leading order in the three-body dynamics as a tidal field across the spatial scattering domain. Although in this analysis
we perform actual 3-body PN simulations with a nearby SMBH, we first explore analytically how
a tidal field might impact the results from the above section.
\begin{figure} 
\centering
\includegraphics[width=\linewidth]{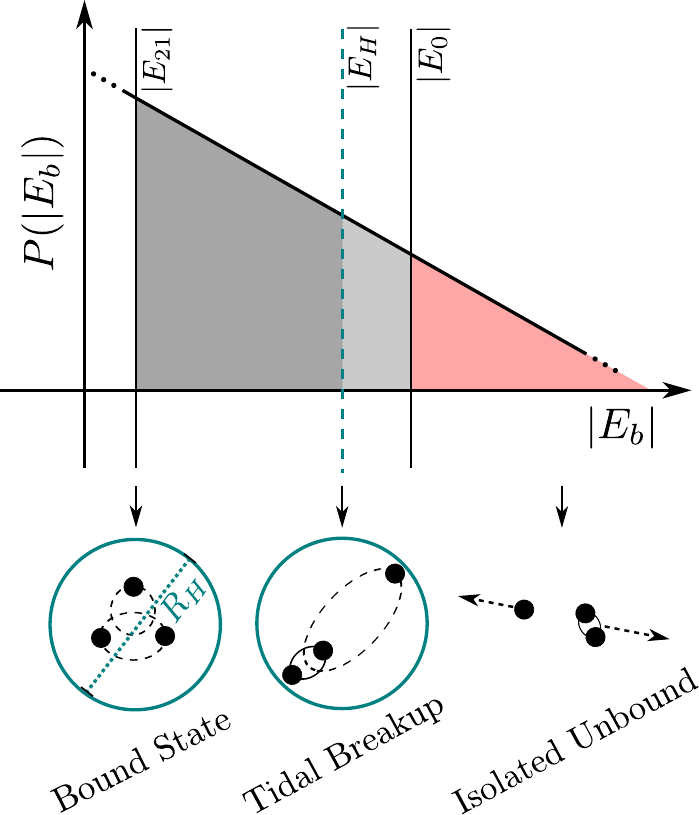}
\caption{Illustration of how the distribution of binary energies, $P(|E_b|)$, as a function of $|E_b|$ relates to different interaction and outcome states of the three-body system
when the interactions take place in a tidal field characterized by a Hill sphere of size, $R_{\rm H}$. As further described in Sec. \ref{sec:tid fields}, $|E_{21}|$ denotes the minimum energy the
binary can have for the three-body system to still be defined as a binary with a bound single, i.e. as a `2+1' state. $|E_{H}|$ is the IMS binary energy at which the SMA of the binary+single,
$a_{bs}$, equals the Hills sphere, $R_{\rm H}$. $|E_{0}|$ is the initial energy of the IMS binary, and defines the limit of escape in the isolated binary-single interaction problem. The ratio between the different areas enclosed by the energy limits is related, for a given formed IMS binary, to the outcome probabilities. This is illustrated in Eq. \ref{eq:N_IMS} and Eq. \ref{eq:NH_IMS}. 
\label{fig:PeB}}
\end{figure}
For this we imagine the interaction to be taking place in a tidal field that we assume can be represented by the size of the
Hill radius,
\begin{equation} \label{eq:Rhill} 
    R_{\rm H} = R_{\rm CM} \left(\frac{m}{M_{\rm BH}} \right)^{1/3}, 
\end{equation}
where $R_{\rm CM}$ is the distance of the binary center-of-mass (CM) from the central black-hole (as later described in  Sec. \ref{sec:config}), $m$ is the stellar-origin BH mass and $M_{\rm BH}$ is the SMBH mass (as we consider equal mass stellar-origin BHs). By definition, outside the sphere the tidal forces from the perturber separate the interacting objects, which then lead to a
termination of the scattering. An example of such
a sphere, or distance, is also shown in Fig. \ref{fig:BSscattering} (see figure caption).
Assuming the interaction inside the Hill sphere is unaffected by the perturber, the effect from a tidal field on the three-body scattering is therefore to terminate
the interaction before it would reach its natural end-state (defined to be the outcome in the isolated case).
If we translate this into the quantities we use in the theoretical model from Sec. \ref{sec:Formation of Eccentric Mergers}, this means that
a tidal field reduces the number of IMS binaries $\mathcal{N}$, which would therefore lead to a reduction in the probability of both 3-body
mergers and eccentric GW mergers.
In other words, there are less chances, i.e. $\mathcal{N}$, to undergo a 3-body merger when tides can terminate the binary-single interaction.
In the following, we estimate how tides entering in this Hill sphere picture affects the number $\mathcal{N}$, and thereby the resultant merger probabilities.

The dynamics giving rise to the three-body system temporarily splitting into a successive series of $\mathcal{N}$ IMS binaries and bound singles,
relate to the energy exchange between the binary and the incoming single. In the initial configuration the incoming single is unbound with respect to the binary, but after the first
close interaction it most likely happens that the single delivers some of its dynamical energy to the binary, which then expands slightly as a response. If the energy exchange
is larger than the initial energy between the single and the binary, the single gets temporally bound to the binary.
This bound state is what we call an IMS. The IMS can essentially be considered as a regular `binary', with one object being the `IMS binary' (with mass $= 2m$) and the other the `bound single' (with mass $= 1m$),
with a SMA $a_{bs}$ that is given by \citep{Samsing14},
\begin{equation}
a_{bs} = \frac{Gm^2}{|E_0|-|E_b|}
\label{eq:abs}
\end{equation}
where $|E_0|$ and $|E_b|$ here denote the initial energy of the three-body system and the energy of the IMS binary, respectively.
Relating this to the Hill sphere boundary, we now note that the interaction terminates if $a_{bs} > R_{H}$. To calculate the correction from the Hill sphere, we therefore have to
calculate the probability that the 3-body system splits into an IMS with $a_{bs} > R_{H}$. For this, we first use that the total initial energy of the three-body system in the considered hard binary limit,
is about the binding energy of the initial binary, i.e. $|E_0| \approx Gm^2/(2a_0)$. In this limit the energy of a given IMS binary relates to the binary+single SMA $a_{bs}$ by
\begin{equation}
|E_{b}| = |E_0| \left( 1- \frac{2a_0}{a_{bs}}\right),
\end{equation}
where we have used Eq. \ref{eq:abs}.
By now setting $a_{bs} = R_{H}$, we find
\begin{equation}
|E_{H}| = |E_0| \left( 1- \frac{2a_0}{R_H}\right) ,
\label{eq:EH}
\end{equation}
which defines the IMS binary energy at which the system splits into a binary+single state with $a_{bs} = R_{\rm H}$.
To relate this characteristic energy threshold to a probability, we first
have to consider the distribution of $|E_b|$ for the assembly of each IMS binary.
For this we make use of the theory presented in \citet{2006tbp..book.....V, 2019Natur.576..406S}, which through arguments related to statistical mechanics derives
that in the co-planar case the binary energy $|E_b|$ approximately
follows the distribution,
\begin{equation}
P(|E_b|) \propto |E_b|^{-3}. 
\label{eq:Peb}
\end{equation}
To get a better understanding of how these energies relate to different three-body outcomes,
we now consider Fig. \ref{fig:PeB}, which shows $P(|E_b|)$, with three characteristic limits. 
Generally, the more compact the binary in the IMS is, i.e. the larger $|E_b|$ is,
the larger the orbit of the single with respect to the binary is. As shown in the illustrations at the bottom of the figure, this exchange of energy
gives rise to three distinct states and outcomes:
{\it (1)} { \it `Bound State':} Here the energy of the binary $|E_b|$ is so low that the single is on an orbit that has a comparable SMA, $a_{bs}$, to the SMA of the binary, $a_{b}$.
The limit $|E_{21}|$ therefore denotes the minimum energy the binary can have for the system to be described as an IMS.
{\it (2)} {\it `Tidal Breakup':} At the critical energy $|E_H|$, the SMA of the bound single with respect to the binary, $a_{bs}$, is similar to the Hill Sphere distance, $R_H$. If $|E_b|$ is greater than this value, the system
experiences a tidal breakup. Therefore, if $|E_{21}|<|E_b|<|E_H|$, then the system continues interacting as an IMS, whereas if $|E_b|>|E_H|$ the interaction terminates.
{\it (3)} {\it `Isolated Unbound':} When there is no tidal field, the system naturally breaks up when the energy of the single with respect to the binary is positive, i.e. when it is being sent out on an unbound orbit
by the binary. In the isolated case this happens when $|E_b|>|E_0|$ (in the hard binary limit).
The ratio between the different areas enclosed by the energy limits ($|E_{21}|, |E_{H}|, |E_{0}|, |E_{\infty}|$),
relates to the outcome probabilities for a given formed IMS binary. We use this picture to calculate how the effect from a tidal field of size $R_{\rm H}$,
as a tidal breakup limits the possibility for forming 3-body- and eccentric mergers by breaking the resonance chain of IMS binaries.

With this formalism we can now estimate the number of IMS, $\mathcal{N}$, with and without a tidal field. In the absence of the tidal field, $\mathcal{N}$ is given by the ratio
between the bound-state area, and the escaper-area,
\begin{align}
	\mathcal{N} & \approx \int_{|E_{21}|}^{|E_0|} x^{-3} dx / \int_{|E_0|}^{\infty} x^{-3} dx\nonumber\\
		   	   &  \approx \left(E_{21}/E_0\right)^{-2},
	\label{eq:N_IMS}
\end{align}
where we have used the distribution from Eq. \ref{eq:Peb}.
In the case of system constrained by a Hill sphere $R_{\rm H}$, the corresponding number of IMS, $\mathcal{N}_H$, is instead given by the ratio:
\begin{align}
	\mathcal{N}_H & \approx \int_{|E_{21}|}^{|E_H|} x^{-3} dx / \int_{|E_H|}^{\infty} x^{-3} dx\nonumber\\
		   	   &  \approx \left(E_{21}/E_H\right)^{-2},
	\label{eq:NH_IMS}
\end{align}
Since the probability for the merger types we consider is $\propto \mathcal{N}$, as seen in e.g. Eq. \ref{eq:P3b}, we can now deduce
that when a system is tidally limited, then the probability for merger during the interaction is reduced by the factor,
\begin{equation}
\frac{\mathcal{N}_H}{\mathcal{N}} \approx \left( \frac{E_H}{E_0} \right)^{2} \approx \left(1 - \frac{2a_0}{R_{\rm H}} \right)^{2},
\end{equation}
where we have used the relation from Eq. \ref{eq:EH}.
From this we can conclude that when the interaction is limited by the Hill sphere, $R_H$, the merger probabilities are corrected in the following way, 
\begin{equation}\label{eq:correction}
P_{3b}(R_{\rm H}) \approx P_{3b} \times \left(1 - \frac{2a_0}{R_{\rm H}} \right)^{2}. 
\end{equation}
If we ask at what $R_{\rm H}/a_0$ the probability has decreased by a factor of $2$, one finds that happens when $R_{\rm H}/a_0 \sim 7 $.
Therefore, if the binary is within a few times the Hill sphere to start out with, then the outcomes
from isolated interactions without the tidal perturber are expected to return probabilities that are accurate within a factor of order of unity. 
With a solid theoretical understanding and models to test, in the section below we treat this problem with PN N-body simulations.

\section{Model and Methods} \label{sec:Methods}

Having described and considered various theoretical aspects of how tides might impact the chaotic interaction of binary-single scatterings and their outcomes, we now turn to
exploring this problem in more detail using PN N-body simulations. 
As the key interest is related to how eccentric mergers are produced at high rates in disc-like environments, where the population of co-planar objects is significant,
we start here by focusing on the co-planar setup. After this we move on to out-of-plane configurations in Sec. \ref{sec:3D}.
In the following sections we present our adopted model, numerical methods, and results from N-body scatterings including a nearby SMBH.

\begin{figure*}
\centering
\includegraphics[width=\linewidth]{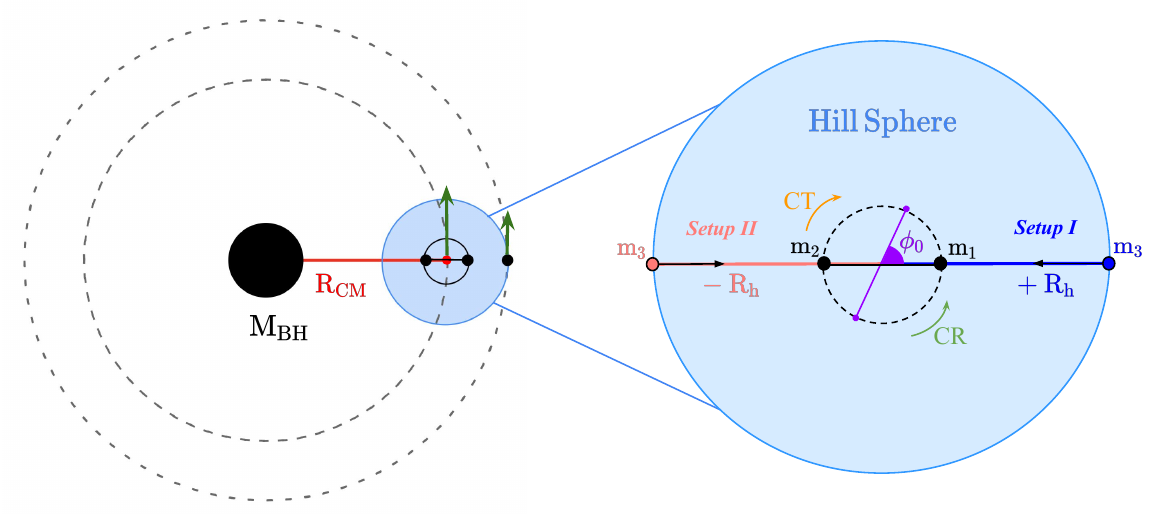}
\caption{Overall (left) and close-up (right) schematic representation of the scatterings configuration. The SMBH is at the center, and the BBH center-of-mass is placed at a distance $R_{\rm CM}$ from the SMBH (red line on the left side). The triple BH system (BBH formed by $m_1$ and $m_2$, and incoming single BH $m_3$) is here on a co-planar circular orbit with respect to the SMBH (2D configuration). The blue circle represents the Hill sphere with radius defined in Eq. \ref{eq:Rhill}. We test two different configuration setups. \textit{Setup I} (blue): single BH is approaching the binary from the right; \textit{Setup II} (pink): the single BH is placed closer to the SMBH and approaches therefore the binary from the left. We test both BBH co-rotating (CR) and counter-rotating (CT) with respect to the orbit around the SMBH. For a fixed SMA, we vary the initial binary phase angle indicated by $\phi_0$. 
\label{fig:setup}}
\end{figure*}

\subsection{Initial configuration of binary-single system orbiting the SMBH}\label{sec:config}

In this work we directly expand on the analysis of \citetalias{Samsing22}, most notably on the result that in general, co-planar scatterings result in order-of-magnitude more eccentric BBH mergers compared to the normal isotropic case taking place in e.g. stellar clusters \citep{Samsing18}. However, the results in
\citetalias{Samsing22} were based on scatterings without a nearby SMBH, and did as well include outcomes from a range of impact parameters that do not necessarily reflect the conditions expected in an AGN disc.
Motivated by this, we here extend their analysis to now also include a nearby SMBH and ICs that likely are closer to the ones taking place in AGN accretion-disc-like  environments.
More specifically, we here perform scattering experiments between a binary black hole (BBH) and a single black hole while they orbit a
SMBH with mass $M_{\rm BH}$ on a circular orbit with their center of mass (CM) placed at a radius $R_{\rm CM}$, as further discussed and shown in Fig. \ref{fig:setup}.
Since we expect embedded orbiters to move subsonically, we set the initial eccentricity of the binary system $e_{b}=0$,
as we imagine that the binary in the subsonic regime has been circularized through (previous) interactions with gas \citep[e.g.][]{ONeill24}.
The BBH and the single BH are placed at the beginning of each simulation on their own circular orbits around the SMBH. The single
BH is put at the gravitational sphere of influence of the triple with respect to the SMBH, which is set by the Hill radius shown
by blue circle in Fig. \ref{fig:setup} and its size is set by Eq. \ref{eq:Rhill}.
This estimate for the tidal influence of the central black-hole on the three-body system is only used for setting up the initial conditions, as the evolution of
the few-body systems naturally includes effects from the SMBH through the N-body solver.

The illustration to the right in Fig. \ref{fig:setup} shows a close-up of the initial configuration.
For each scattering experiment, we vary the initial phase of the binary angle $\phi_0$.
In addition, we explore co-rotating (CR, green arrow) and counter-rotating (CT, orange arrow) configurations, both of which are believed to take place
in AGN disc environments \citep{2021ApJ...908L..27S}. The definition of CR and CT is naturally set by the direction of motion of the triple system with respect to the SMBH. With this, we explore
two different setups as shown in the right panel of Fig. \ref{fig:setup}: 
 \begin{itemize}
     \item {\it Setup I} (blue): The single BH is initially placed at the outer boundary of the Hill sphere w.r.t. the SMBH, so that its
     distance from the SMBH is $R_{\rm CM} + R_{\rm H}$.
     \item {\it Setup II} (pink): The single BH is initially placed at the inner boundary of the Hill sphere w.r.t. the SMBH, so that its
     distance from the SMBH instead is $R_{\rm CM} - R_{\rm H}$.
 \end{itemize}
The most likely initial configuration depends on the astrophysical settings, e.g. if the binary and the single meet due to
different migration speeds inside the disc, {\it Setup II} is the most likely as the heavier BBH `sweeps up' the lighter single BH as they both migrate through the disc.
On the other hand, if the interactions are taking place in a migration trap, {\it Setup I} is the most likely, as the single BH approaches the BBH from the outer
parts \citep{2019ApJ...878...85S}. If one allows for significantly different masses, both {\it Setup I} and {\it Setup II} can take place throughout the disc. It is important to note that as we vary the position of the incoming single BH, the relative orbital velocity between the BBH and the BH is varied as well (indicated by the green arrows on the left side of Fig. \ref{fig:setup}).
With these more physical realistic ICs, we now explore the possible differences between the results from \citetalias{Samsing22} and the results
from the scatterings.

\subsection{Merger classification} \label{sec:merger}

The end-state of few-body scatterings can generally be classified and divided into a range of distinct outcomes depending on what
environments the scatterings are taking place in, on the
probed observables, and on how finite sizes as well
as energy dissipative terms (GW radiation, tides, gas, etc.) are included \citep[e.g.][]{Samsing17}. In this work we only report results for BBH mergers that
have a notable eccentricity, $e$, at frequency $f_{\rm gw}$ as these carry unique information about the nature of the formation channel.
In addition, when focusing on eccentric
BBH mergers in the LVK bands, the merger time is so short that we do not have to complicate the study by introducing and working
with several outcome types. For example, if we considered eccentric LISA sources instead, the relevant population would be the BBHs that do not undergo
a 3-body merger but instead are left unbound after the interaction with a notable eccentricity \citep{2018MNRAS.tmp.2223S,2018MNRAS.481.4775D}. However, in the AGN disc, tracking this
population is very difficult, as the evolution of such binaries after the interaction is likely to either be greatly
affected by the gas, or by other objects embedded in the disc. In stellar clusters this is in contrast relatively easy as was demonstrated in \citet{Samsing18}.
Focusing on eccentric LIGO/Virgo/KAGRA sources in the adopted AGN setup is more robust, as they form and merge promptly. 
For these reasons, in this paper we mainly focus on the prompt BBH mergers that form in our
simulations and appear with notable eccentricity (>0.1) in the LVK bands ($\sim 10\ \rm Hz$).

\subsection{Numerical methods and PN corrections} 

We perform simulations with the same extensively tested N-body code as in \citet{Samsing14,Samsing17,Samsing18,Samsing22}, which adopts a solver based on \texttt{LSODE} with a relative and absolute error set to $10^{-12}$ and an adaptive time-step with a scale based on the relative position, velocity and acceleration of the interacting bodies following standard procedures. 
The code includes GR effects by the use of the
PN formalism \citep{Blanchet06, Blanchet14}. In this formalism the effects from GR are added to the Newtonian acceleration as an expansion in $v/c$, 
\begin{equation}
    \boldsymbol{a} = \boldsymbol{a}_0 + c^{-2} \boldsymbol{a}_2 + c^{-4} \boldsymbol{a}_4 + c^{-5} \boldsymbol{a}_5 + \mathscr{O} (c^{-6}),
\end{equation}
where $\boldsymbol{a}_0$ denotes the newtonian acceleration (0PN order), $c^{-2} \boldsymbol{a}_2 + c^{-4} \boldsymbol{a}_4$ (1PN + 2PN order) are energy conserving
and lead to orbital precession \citep{Blanchet06}, and finally $\boldsymbol{a}_5$ is the dissipative term that in this formalism is the leading order term describing the energy
and momentum loss due to GW radiation. When doing an orbit average, the effect from this term is essentially equivalent to the well-known \citet{Peters64} formulae.
In this work we do not include the conservative 1PN, 2PN corrections, as these terms are essential for evolving hierarchical systems where the interactions are taking place over many orbits,
but have been shown not to play a significant role in the type of chaotic scatterings we are exploring here.
We focus instead on the leading effect from GW energy dissipation for assembling BBH mergers, i.e.
the terms we include in the N-body code are $\boldsymbol{a} = \boldsymbol{a}_0 + c^{-5} \boldsymbol{a}_5$. This PN-acceleration is applied pairwise between
any two objects `1' and `2' such that the acceleration terms can be written as,
\begin{equation}
    \boldsymbol{a}_0 = -\frac{Gm_2}{r_{12}^2} \hat{\boldsymbol{r}}_{12},
\end{equation}
and
\begin{align}
    & \boldsymbol{a}_5 = \frac{4}{5}\frac{G^2 m_1 m_2}{r^3_{12}} \Bigl[\left( \frac{2Gm_1}{r_{12}}-\frac{8Gm_2}{r_{12}}-v^2_{12}\right) \boldsymbol{v}_{12} \\
    & \qquad + (\boldsymbol{\hat{r}}_{12}\boldsymbol{v}_{12}) 
     \left( \frac{52Gm_2}{3r_{12}} - \frac{6Gm_1}{r_{12}} + 3v^2_{12} \right) \boldsymbol{\hat{r}}_{12} \Bigr] \notag
\end{align}
where bold symbols indicate vectors. Using the same notation as in \citet{Blanchet06}, the separation vector is defined here as
$\boldsymbol{r}_{12}= \boldsymbol{r}_{1}-\boldsymbol{r}_{2}$ with
corresponding relative velocity vector $\boldsymbol{v}_{12}$, and unit vector $\hat{\boldsymbol{r}}_{12}$ given by $\boldsymbol{r}_{12}/r_{12}$.
We note that PN terms have been derived up to higher orders; however, it is not clear how the series converges, and for stability we restrict ourselves
not to include higher order terms.

\section{Results}

We here present the main results on the formation of eccentric BBH mergers forming as a consequence of binary-single interactions
taking place near an SMBH in a disc-like setup. Below we start by showing a few illustrative cases where the tidal field from the SMBH clearly plays a role
in the interaction and outcome. We then show the probability for forming eccentric mergers as a function of the BBH semi-major axis (SMA), relative to the
Hill sphere of the SMBH, for both co-rotating and counter-rotating interactions. These results are compared to the scatterings by \citetalias{Samsing22}. \footnote{We note that in \citet{Trani23},
a large fraction of eccentric mergers were also found in their setup with a SMBH, but the ICs and underlying physical setup (inclusion of stellar-disc velocity dispersion) are different from ours, which makes
it difficult to compare side-by-side.}. The new features we find are then explored by considering the phase-space, or topology, of the scatterings. Finally,
we show results for out-of-plane scatterings, as the binary-single orbital inclination is believed to be critical in the production of eccentric mergers as shown in \citetalias{Samsing22}.

\subsection{Escape vs. capture binary-single interactions}\label{sec:single}

\begin{figure}
\centering
\includegraphics[width=\linewidth]{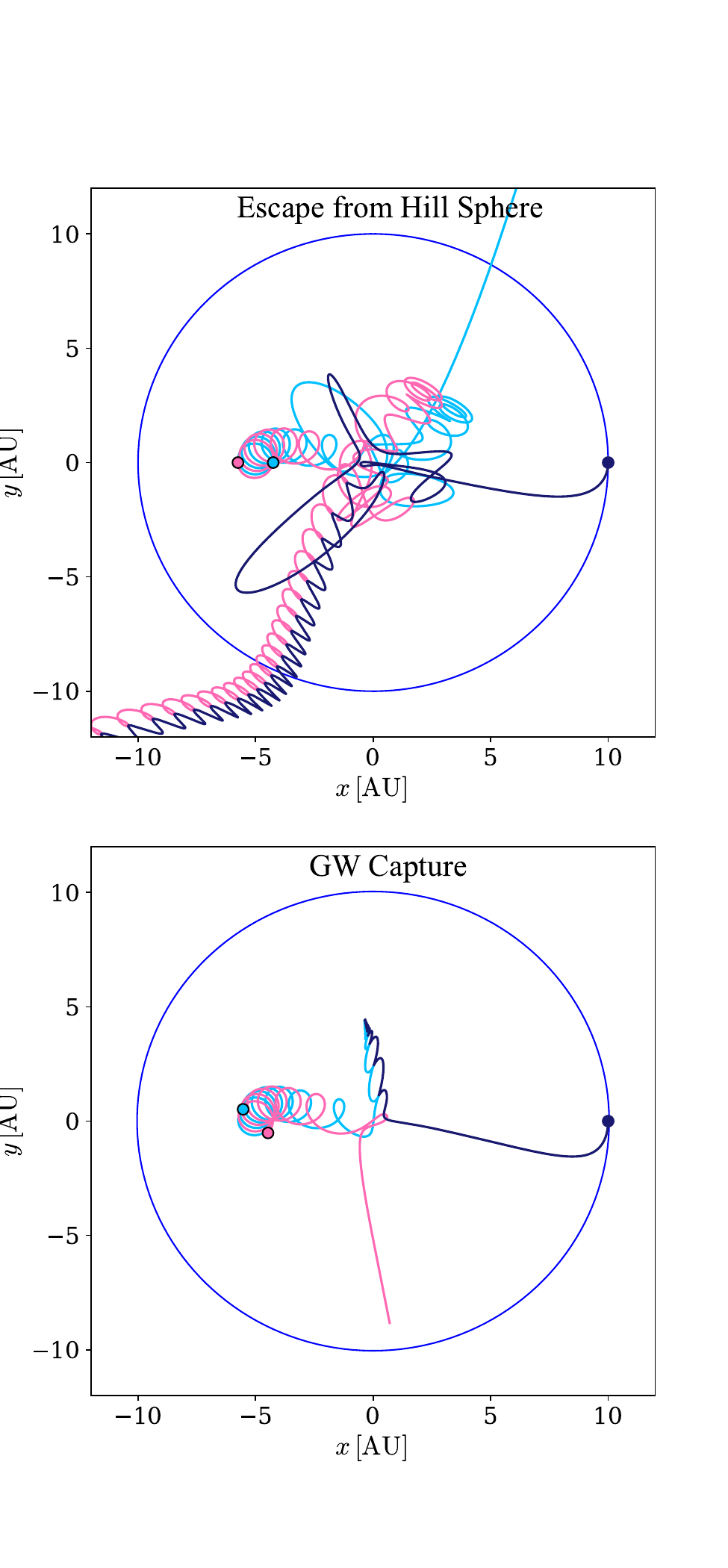}
\caption{Examples of equal-mass binary-single interaction under the influence of the SMBH resulting in an escape from the Hill sphere (top) and a 3-body GW merger (bottom) for \textit{Setup I}.
The dots indicate the initial positions. The encounter is plotted with respect to the three-body center-of-mass frame, where the single BH is placed at the Hill sphere indicated by the blue circle
with a radius set to be 10 AU. The initial binary SMA is 1 AU for both experiments. The initial binary phase in the top panel is set to $0^{\circ}$ and $135^{\circ}$ for the bottom panel,
indicating that with fixed initial SMA and $R_{\rm CM}$, varying the BBH phase can easily change the scattering outcome.
\label{fig:combo}}
\end{figure}

Fig. \ref{fig:combo} shows two different examples of three-body scatterings for \textit{Setup I}, where the single BH is approaching the BBH from the outside, as earlier
described in Fig. \ref{fig:setup}. Both scatterings are plotted with respect to the binary+single CM. The top panel displays an interaction that is terminated before
its natural outcome as a result of the tidal field from the SMBH, which breaks up the binary as the interaction here leads to an IMS that extents outside of the Hill
sphere (blue circle). For this scattering, the initial binary semi-major axis is 1 AU and the distance from the SMBH is chosen such that the size of the Hill radius
is 10 AU. The initial binary phase angle $\phi_0$ is 0$^{\circ}$.
For the experiment in the bottom panel, the SMA and Hill radius are kept the same as in the left, but the binary angle is instead 135$^{\circ}$.
As shown, this variation in the binary phase leads to a completely different outcome, namely the merger between two of the three
objects (pink, dark-blue) while the single BH is still bound, i.e. the outcome we denote a `3-body merger'. This outcome is not possible in a purely Newtonian code,
as the inspiral of the two objects undergoing the merger is entirely driven by the radiation arising from the 2.5PN term.
Comparing the two shown scatterings, we clearly see that some interactions almost promptly undergo a merger (bottom panel), whereas some
enter the resonating state as we saw in Fig. \ref{fig:BSscattering}. As described in Sec. \ref{sec:tid fields}, when the system enters these chaotic states the tidal field from the SMBH is likely to split it apart before e.g. a 3-body merger is formed. If gas is present one can imagine the drag forces acting continuously on the bottom panel interaction, which would gradually transfer energy out of the entire system. Such processes could essentially protect the system from undergoing a tidal breakup as it would harden over time. In this simple picture, the gas in the accretion disc could therefore
help increase the number of 3-body mergers that otherwise could have terminated earlier by the SMBH tidal field.

\subsection{Probability of eccentric black-hole mergers}\label{sec:class}

\begin{figure}
\centering
\includegraphics[width=1.1\columnwidth]{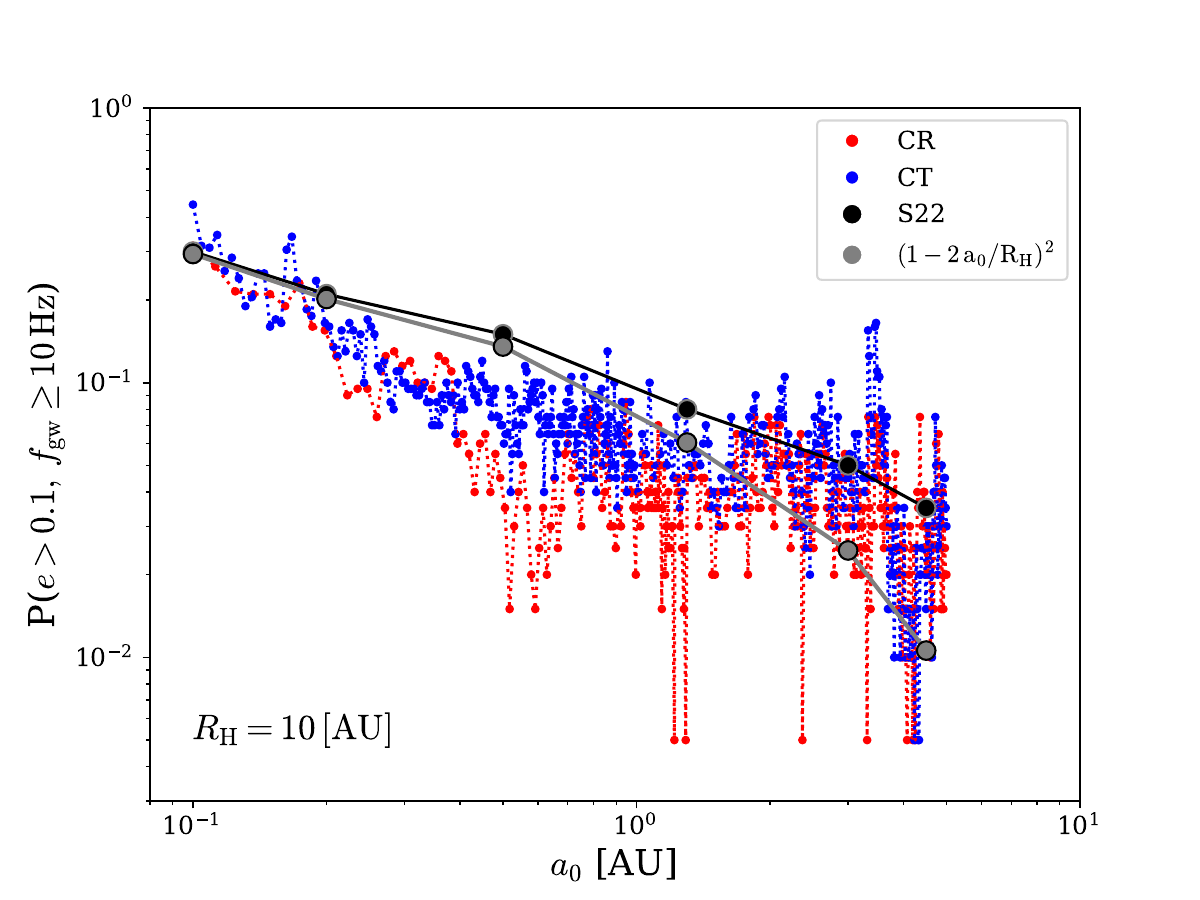}
\caption{
Probability for forming a GW merger with $e>0.1$ at peak frequency $f_{\rm gw} \geq$ 10 Hz, for co-rotating (red) and counter-rotating (blue) BBHs in setup I, for $m = 100 M_{\odot}$ triple
BH system. The large black dots are the replicated results from \citetalias{Samsing22}. The grey black dots represent the correction factor due to the tidal field
according to eq. \ref{eq:correction}. At lower semi-major axis (up to $\approx$ 0.3 AU) the trend of the probability follows approximately a power law, that then breaks after which the
curve becomes more fluctuating as the size of the SMA approaches the size of $R_{\rm H}$. Further analysis in Sec \ref{sec:orbres}. 
\label{fig:linefit}}
\end{figure}

Fig. \ref{fig:linefit} shows the probability for the considered scatterings to result in an eccentric merger ($e>0.1$) with a peak frequency $f_{\rm gw} \geq 10$ Hz (LVK band), as a function of the semi-major axis of the target BBH. For this figure we have assumed $m = 100 \, M_{\odot}$, $M_{\rm BH} = 10^{8} \, M_{\odot}$, and that the
single encounters the binary from a distance equal to a Hills sphere $R_{\rm H}$ = 10 AU, which corresponds to $R_{\rm CM} = 10^3$ AU. For each point in the figure
we vary the phase angle $\phi_0$ from 0 to 2$\pi$ in $N$ steps, where $N = 2 \cdot 10^5$ for the results presented here.
The probability is estimated by taking the total number of outcomes resulting in an eccentric inspiral ($e>0.1$ at $f_{\rm gw} \geq 10$ Hz)
to the total number of scatterings per SMA ($N = 2 \cdot 10^5$).

Generally, we do not evolve each binary until merger, therefore to classify which BBHs are falling into the category of
having an $e>0.1$ at $f_{\rm gw} \geq 10$ Hz, we take the binary output from the N-body code, which essentially corresponds to the end-state binary SMA and eccentricity,
and evolve it using \citet{Peters64} equations. Specifically, first we estimate what pericenter distance, $r_f$, corresponds to a given GW peak-frequency, $f_{\rm gw }$, using Eq. \ref{eq:rf}.
After this, we take the binary SMA and eccentricity outputted from the code, here denoted by $a_{\rm out}$ and $e_{\rm out}$, respectively,
and derive the constant $c_{\rm out}$,
\begin{equation}\label{eq:c0}
    c_{\rm out} = \frac{a_{\rm out}(1-e_{\rm out}^2)}{e_{\rm out}^{12/19}\left[1+\frac{121}{304} e_{\rm out}^2\right]^{870/2299}}.
\end{equation}
We then solve for the eccentricity at which the binary reaches a pericenter distance of $r_f$, using the relation,
\begin{equation}
    r_{f} = \frac{ c_{\rm out} \, e^{12/19}}{1+e} \left[1+\frac{121}{304} e^2\right]^{870/2299}.
    \label{eq:rp}
\end{equation} 
For the sources that form with a GW peak frequency above the imposed threshold, e.g. above $10$ Hz, we label these as eccentric mergers as well \citepalias{Samsing22}.
Theoretically, these mergers belong to the category of being eccentric in the observable band, whereas from an evolutionary point of view, they could have very different
kinds of GW waveforms. For example, mergers with extremely high eccentricity show up as burst sources with a GW signal similar to e.g. GW190521 \citep{2020CQGra..37e5002A}, whereas less eccentric mergers
gradually inspiral throughout the observable band.
For the results in Fig. \ref{fig:linefit}, we map both co-rotating (red) and counter-rotating (blue) binaries from \textit{Setup I}.
As one of the main goals with our analysis is to explore if the findings from \citetalias{Samsing22}, stating eccentric mergers are produced at
exceptionally high probability in AGN-environments, also holds for more realistic ICs and with the inclusion of a SMBH, we here also show data from \citetalias{Samsing22}
in black. From this we can now conclude that in terms of the overall scaling and dependency on SMA, the results with a SMBH and with the single approaching from the
Hill sphere is similar to the results found in \citetalias{Samsing22}, which were based on isolated 3-body scatterings (and no Hill's boundary).
This especially implies that the co-planar geometry that likely
is facilitated by the AGN-disc, is still highly effective in creating eccentric GW sources, despite the influence from the nearby SMBH. 
Looking closer at the scattering results, we do see some unique features in the data when the SMA
of the BBH approaches the Hill sphere $R_{\rm H}$, i.e. in this case when $a_0$ approaches $10$ AU. In general, the more we increase the semi-major axis, i.e.
the binary size becomes comparable to the Hill radius, the more the probability trend becomes less linear and much more fluctuating. This is not due to numerical noise, but a
real feature of the scatterings.
There are even regions in which the probability increases with increasing binary size (e.g. after 1 AU, where $\rm a_0 / \rm R_{\rm H}=0.1$), to then start decreasing again (after 3 AU).
Although a direct comparison cannot be made, \citet{Boekholt23} also found clear (fractal) structures in their two-body Jacobi captures, which also
results from keeping track and varying the ICs in a systematic way as we are also doing, but here in the three-body problem.

\begin{figure}
\centering
\includegraphics[width=1.1\columnwidth]{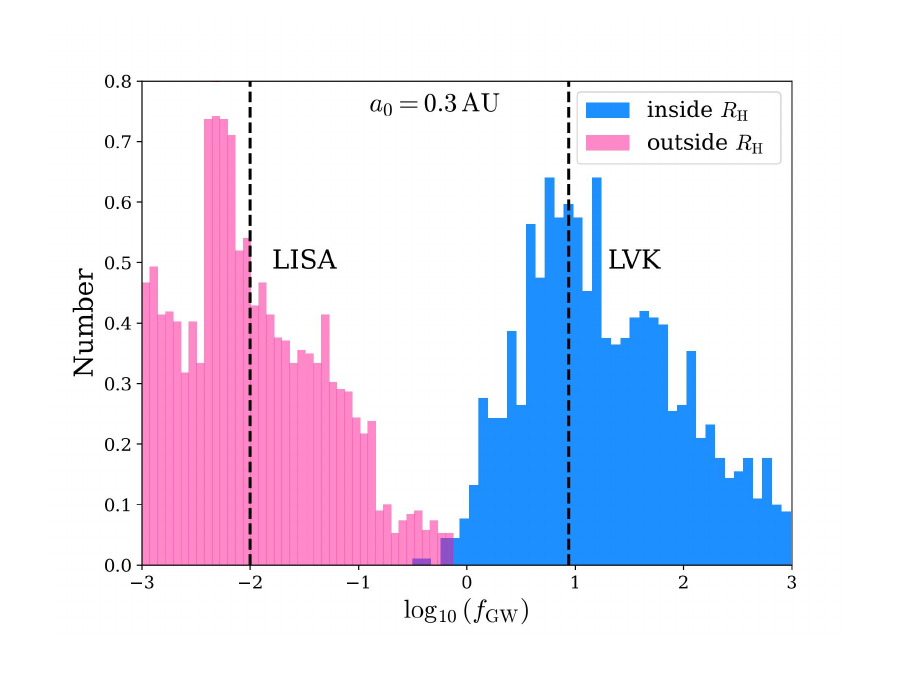}
\caption{
Distribution of GW peak frequency of binaries formed through binary-single interactions near a SMBH evaluated at the time of assembly (see Sec. \ref{sec:class}).
The initial SMA is $a_0 = 0.3 AU$.
The blue distribution represents the binaries that merged inside the Hill sphere (3-body mergers), while the pink distribution indicates binaries that survived the
interaction. Over-plotted are the LISA and LVK peak frequency bands (dashed vertical black lines), showing that a large fraction of sources enter both LISA and LVK bands at formation. 
\label{fig:fgw_dist}}
\end{figure}

Finally, Fig. \ref{fig:fgw_dist} shows the actual distribution of GW peak frequency returned from the scatterings at the time we stop the simulation, where the stopping criterion corresponds to either a GW capture inside the Hill sphere, or to the interruption of the simulation if the outcome is an escaper, where all three objects have left the Hill sphere (same as the two outcomes shown
in Fig. \ref{fig:combo}).  
The chosen value for the SMA is 0.3 AU, for which $\phi_0$ is varied $10^4$ times. The {\it blue} distribution indicates the binaries that underwent a GW capture inside the Hill sphere, while the {\it pink} distribution represents the population of binaries that survived the interaction. Both eccentric LISA and LVK mergers are forming,
which would play a major role when multi-band GW observations become possible with LISA, LVK, and possible 3G instruments operating simultaneously.
Starting with the LVK sources, we see that a significant fraction of mergers are assembled everywhere
from 10 to $10^3$ Hz, implying
that many of the mergers we are finding are closer to burst sources or direct plunge sources than to classical eccentric mergers that almost adiabatically evolve from high eccentricity to circular inspiral.
Burst and plunge searches are therefore equally important to perform as eccentric searches for probing the contribution of disc-like environments. 
We note that such searches have been done \citep{2020CQGra..37e5002A,2016PhRvD..93d2004K} with no clear detections yet, except maybe for GW190521 \citep{GW19b}. 

\subsection{Effects of periodic orbital encounters}\label{sec:orbres}

\begin{figure*}
\centering
\includegraphics[width=\linewidth]{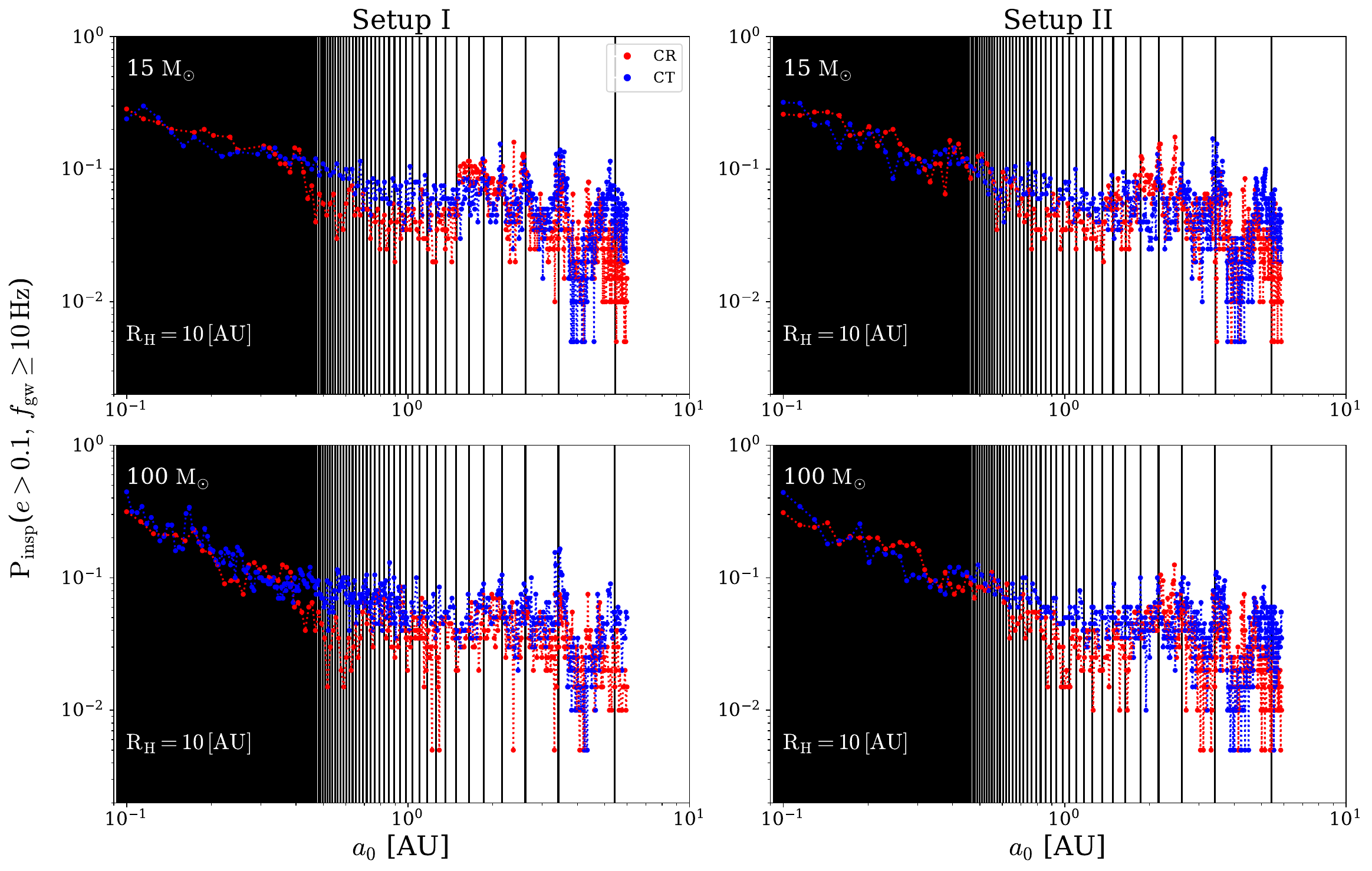}
\caption{Probability of mergers at $e>0.1$ and $f_{\rm gw} > 10 \, \rm Hz$ as a function of initial inner binary semi-major axis SMA (similar to Fig. \ref{fig:linefit}). The left column displays results for Setup I for a 15 $M_{\odot}$ (top) and 100 $M_{\odot}$ (bottom) equal mass triple system, while the right column corresponds to results with \textit{Setup II} for the same BH masses. As in Fig. \ref{fig:linefit}, the binary SMA ranges from values of 0.1 to 6 AU both CR(red) and CT(blue). The black vertical lines represent solutions to Eq. \ref{eq:resma}, where $n$ equals an integer number of half-orbits.
As the size of the SMA approaches $R_{\rm H}$, the fluctuating regions become more spread out and the trend of the probability becomes less linear and more fluctuating, as described in Sec. \ref{sec:orbres}.
\label{fig:prob}}
\end{figure*}

To explore the nature of the fluctuating scattering results seen in Fig. \ref{fig:linefit}, we first consider Fig. \ref{fig:prob}, which further
expands on the results by
showing the probability for an eccentric merger for two different BH masses (15 $M_{\odot}$, top panels and 100 $M_{\odot}$, bottom panels) and for
\textit{Setup I} (left panels) and \textit{Setup II} (right panels). All four configurations show similar behavior to the one described in Fig. \ref{fig:linefit}, i.e. at lower SMA,
the probability mostly follows a linear trend, while it becomes increasingly more fluctuating as we widen the BBH semi-major axis. The general reason is that our
setup, where the single approaches the BBH from the Hill sphere, is so constrained that we are seeing real features in the underlying phase-space
or topology of the three-body system \citep{Samsing2018}. Before studying the topology of the scatterings, we first
explore analytically if we can estimate when the scatterings go from being a near smooth power-law, to greatly varying.
For this, we note that the scatterings most likely have some underlying periodic features, as we keep the initial
distance between the binary and single fixed ($R_{\rm H}$ does not change with the BBH SMA) while changing the SMA
of the target BBH. This setup gives rise to periodic interactions, where the single falls from the Hill sphere
to interact with the BBH after the BBH has undergone $n$ half-revolutions as the
SMA decreases. In other words, we might expect periodic or repeating features when the following condition is met for integer $n$,
\begin{equation}
    n = \frac{T_{\rm int}}{T_{\rm BBH}},
    \label{eq:def_n}
\end{equation}
where $T_{\rm int}$ is the time it takes the single to reach the BBH (interaction time), and $T_{\rm BBH}$
is the orbital time of the BBH. Starting with $T_{\rm int}$, the natural timescale associated with this interaction time,  scales with the parameters of the problem as,
\begin{equation}
    T_{\rm int} \sim \sqrt{\frac{R_{\rm H}^3}{Gm}},
\end{equation}
where for this analysis we assume the BHs to have equal mass, $m$. Correspondingly, the timescale associated with the BBH is simply
\begin{equation}
    T_{\rm BBH} \sim \sqrt{\frac{a_0^3}{Gm}}.
\end{equation}
By using Eq. \ref{eq:def_n}, we can now isolate for the BBH SMA that corresponds to integer values of $n$,
\begin{equation}\label{eq:resma}
     a_n \propto R_{\rm H} \, n^{-2/3}.
\end{equation}
The values of $a_n$ for integer values of $n$ is plotted over the probability distributions shown in Fig. \ref{fig:prob} with {\it black} vertical solid lines.
The vertical lines line up well across the different combinations of ICs,
with the fluctuating and repeating features in the probability curves. Furthermore, because of the scaling
from Eq. \ref{eq:resma}, the BBH takes more and more turns, and the spacing between the lines becomes smaller and smaller, with decreasing SMA,
which explains why in this limit the probability curves have less features and more closely follows a simple power-law. The more the BBH turns and the higher line density for lower binary semi-major axis imply that the system increasingly loses information about the
exact ICs, in which case it approaches the more chaotic limit that, e.g., was shown in \citetalias{Samsing22} to follow a smoother curve. Correspondingly, when the SMA
increases towards $R_{\rm H}$, the impact from the exact ICs is important and the underlying phase-space features start to show up as features in the probability
curves. These observations and the theoretical model from Sec. \ref{sec:tid fields} are all complementary, and clearly indicate that, at least for this case, the largest
visible effect in the probability when introducing the SMBH is not `tidal breakup', but instead fluctuations from the underlying phase-space. However, it might be that
the real astrophysical environment gives rise to more chaotic ICs, e.g. if the single BH does not slowly approach the BBH from the edge of the Hill sphere, in which
case the effect from tidal breakup could be a more dominant effect.
Continuing to consider results from the scatterings shown in Fig. \ref{fig:prob}, we find that the critical $n$ for which
the system can be considered chaotic is for $n \sim 10$. 

Below, we continue by exploring the scattering phase-space, or topology, in greater detail to further understand the
probability features and where and how the eccentric BBH mergers are forming.

\subsection{Topology of end-states}\label{sec:phase}

\begin{figure*}
\centering
\includegraphics[width=\linewidth]{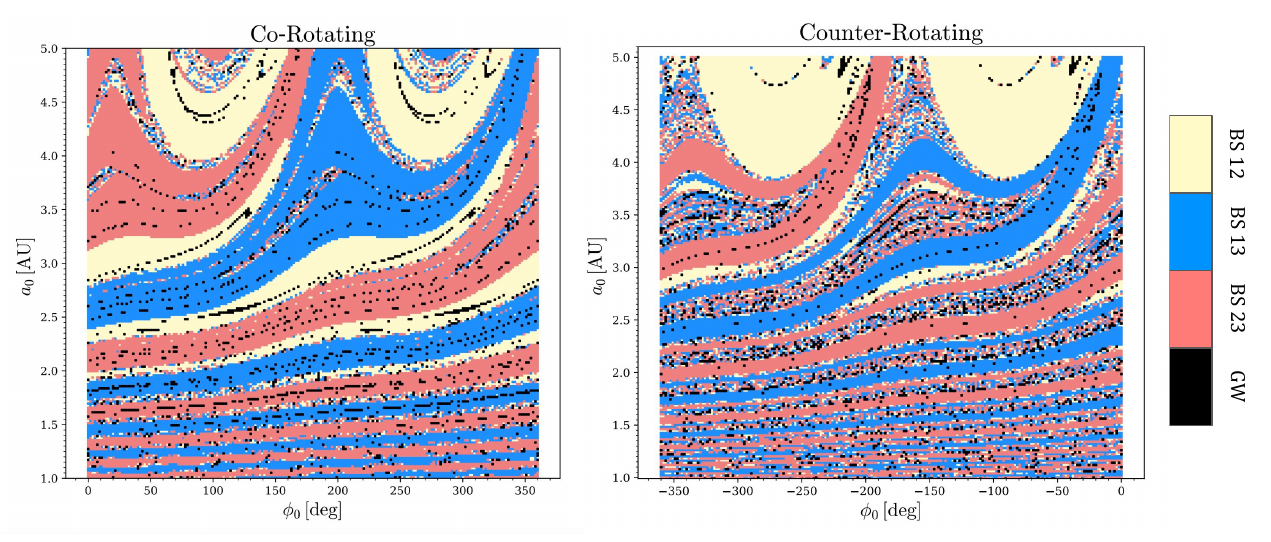}
\caption{Phase-space distribution of a selected region for \textit{Setup I} in the CR(left) and CT(right) case. We fix the impact parameter of the incoming single at the Hill radius and vary the inner BBH SMA and phase angle. This particular section of the phase space is between SMAs of 1 and 5 AU. The plot is color-coded in terms of final outcome as explained in Sec \ref{sec:phase}: Escaper from Hill sphere 12 (yellow), 13 (blue), 23 (pink). The black region corresponds to a 3-body GW merger where two of the three objects merge inside the Hill sphere.
\label{fig:phaseplot}}
\end{figure*}

As we find very similar features for the phase space distribution of scatterings for \textit{Setup I} and \textit{Setup II}, we focus on the comparison of the co-rotating and counter-rotating cases in relation to the probability features, rather than on the comparison of the two configurations. We therefore in this section, only show examples from \textit{Setup I}. Fig. \ref{fig:phaseplot} shows the phase space distribution of outcomes (co-rotating for the left panel, and counter-rotating for the right panel), in terms of the initial binary semi-major axis $a_0$ and phase angle $\phi_0$, as defined in Fig. \ref{fig:setup}. Similarly to \citet{Samsing18a}, the figure is color-coded in terms of final end-state:
\begin{itemize}
    \item BS[ij]: Outcome where the BBH and single BH interact and both escape from the Hill sphere without merging before. The possible combinations can be 12 (yellow), 13 (blue), and 23 (pink), where as shown in Fig. \ref{fig:setup} the initial binary components are labeled by `1' and `2', and the incoming single by `3'.
    \item GW (black): The outcome of the scattering is a 3-body merger, where two of three objects undergo a GW inspiral that leads to a GW merger while all three objects are still within the Hill sphere.
\end{itemize}

\begin{figure*}
\centering
\includegraphics[width=\linewidth]{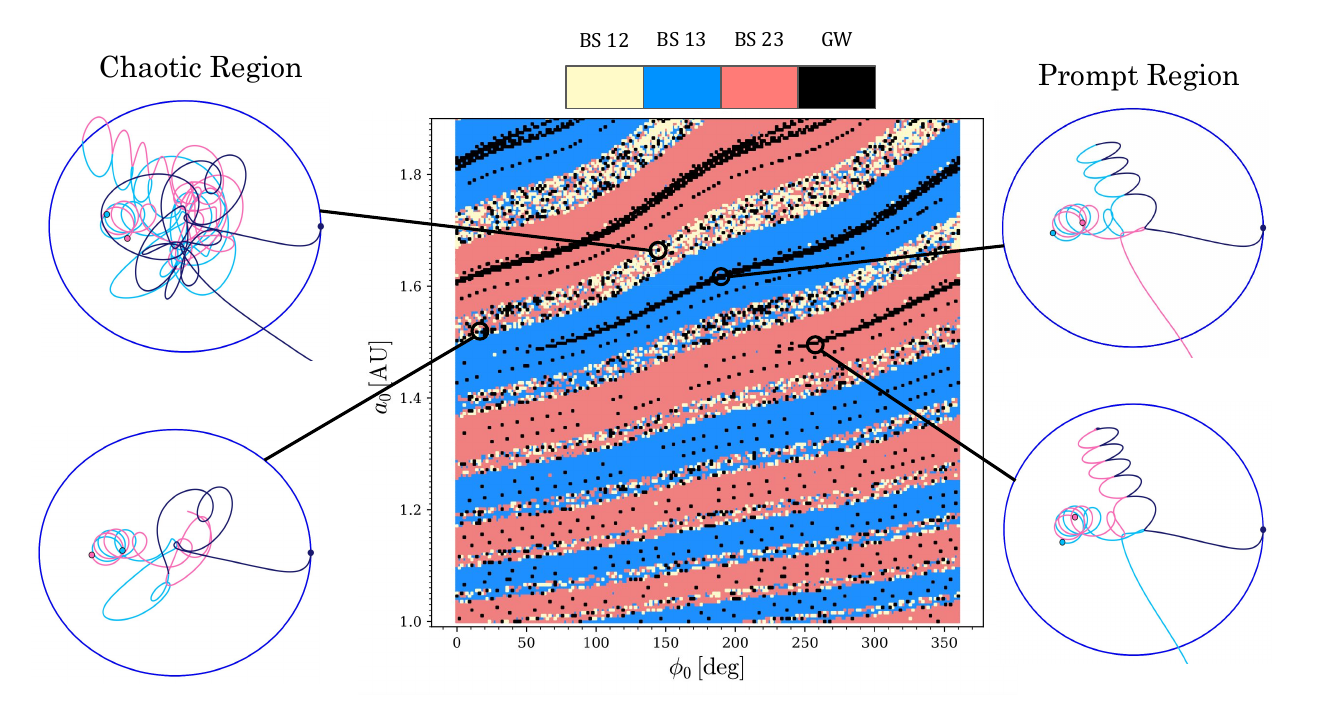}
\caption{Close-up view of parameter space (\textit{Setup I}) with values of initial SMA ranging between 1 and 2 AU, along with four examples of possible interaction scenarios depending on the location in the parameter space. The phase-space appears as divided into bands, alternating between chaotic and prompt regions. The number of interactions for the three-body encounters in the chaotic region is significantly higher than in the prompt regions. 
\label{fig:zoom_1_int}}
\end{figure*}

\begin{figure}
\centering
\includegraphics[width=1.05\linewidth]{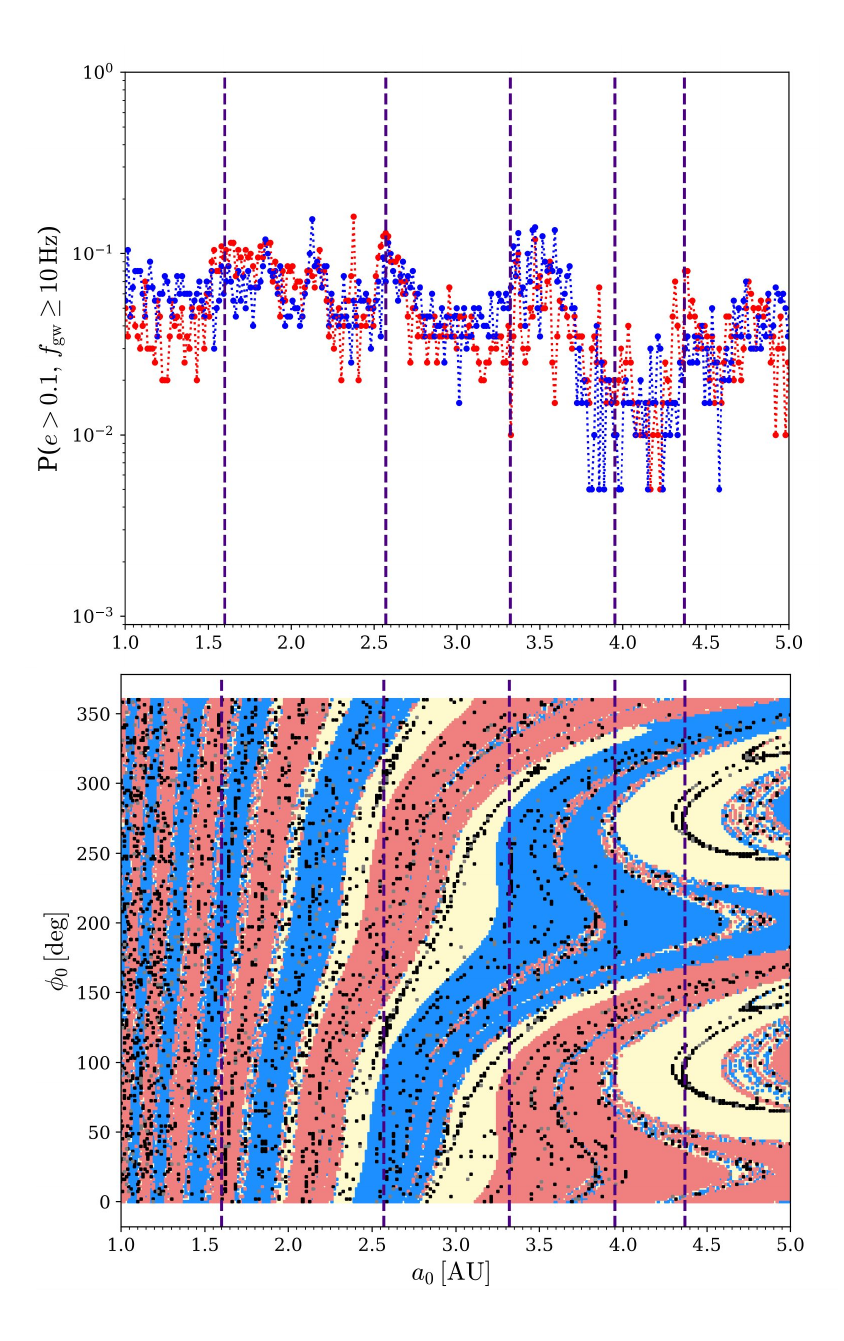}
\caption{Comparison of probability for an eccentric inspiral (top) and the phase space distribution (bottom) for a 15 $M_{\odot}$ equal mass triple system in \textit{Setup I}. The purple vertical dashed lines in both figures are plotted as an indicator of which region of the parameter space the probability corresponds to, showing for which SMAs the probability hits a low or high number of inspirals and how the fluctuation becomes more apparent as the value of $a_0/R_{\rm H}$ increases. 
\label{fig:zoom_2_prob}}
\end{figure}

We only show a portion of the overall analyzed parameter space to look more closely at what happens when the SMA approaches
the size of $R_{\rm H}$. 
From the phase space distribution, a few elements stand out. In terms of $\phi_0$, for both the CR and CT cases the phase-space
structure has a trivial periodicity, where the pattern repeats but in opposite colors after $180^{\circ}$. In terms of the structure along the SMA, we see that as we approach the $R_{\rm H}$ limit, the regions widen and for some angular intervals, e.g. in the CR case (left), there is really only one possible outcome. For example, at $200^{\circ}$ in the CR case, the system can only result in
outcome `BS13' as the binary has no time to undergo significant orbital evolution before it encounters the single, which in turn turns into this restricted outcome space. At lower SMA, the binary can for just small changes in SMA undergo several revolutions, before the single arrives and initiates the interaction (see Eq. \ref{eq:resma}), which give rise to the seen band-like
structure in both the CR and the CT case. 
Another interesting behavior that we further analyse in Fig. \ref{fig:phaseplot}, is that in the CR case (left) there appears to be a significant higher fraction of non-chaotic regions than chaotic ones compared to the CT case (right). Intuitively, this is to be expected from the setup, as the single, which arrives from the Hill sphere, generally encounters the binary from `below' (since the binary in this setup is close to the SMBH than the single), as seen in Fig. \ref{fig:setup}, and therefore initially orbits the binary in the same direction as the CT setup. The objects therefore encounter each other with a lower relative velocity, which generally
leads to a temporary dynamical capture of the single. This effect starts the resonating interaction, which corresponds to the seed to enter a chaotic state. This is in contrast to the CR case, where the single encounters one of the binary
objects with a relatively high velocity, which is more likely to result in a prompt exchange, which might bring the objects out of the Hill sphere after the first encounter. What is not expected is that the resonating interaction and corresponding chaotic regions are so rare in the CR case.  

Furthermore, the phase space figures provide important hints as to how and where the eccentric BBH mergers form,
which is also illustrated in Fig. \ref{fig:zoom_1_int}. Here we highlight mergers formed in the chaotic (left side) and in the prompt regions (right side).
The interactions taking place in the resonating and chaotic regions are much more prone to additional effects such as gas, if present,
and are therefore likely to be greatly affected, as they have much more time to lose energy through drag over the duration of the resonating
state. The opposite can be stated regarding prompt mergers, although they could turn into resonating interactions if enough drag is acting on them before they leave the Hill sphere.

Finally, a more direct comparison between the outcome of probabilities (top) and phase-space (bottom) is shown in Fig. \ref{fig:zoom_2_prob} for a smaller portion of the SMA range (1-5 AU). The vertical dashed lines indicate the corresponding region of the probability in the phase space. Fig. \ref{fig:zoom_2_prob} highlights how the band-like configuration observed in the phase space is connected to the merger probability. As highlighted in Sec. \ref{sec:orbres}, for larger values of $a_0/R_{\rm H}$, the encounters become more controlled as the single BH is starting off closer to the binary. As a result, the regions of orbital resonance shown in \ref{fig:prob} become more spread-out causing large fluctuations in between these regions. Consequently, as the system becomes more sensitive to the initial conditions, the outcome of the scattering (prompt merger or short interaction and escape) strongly depends on the combination of $\phi_0$ and $a_0$.

\begin{figure}
\centering
\includegraphics[width=1.1\columnwidth]{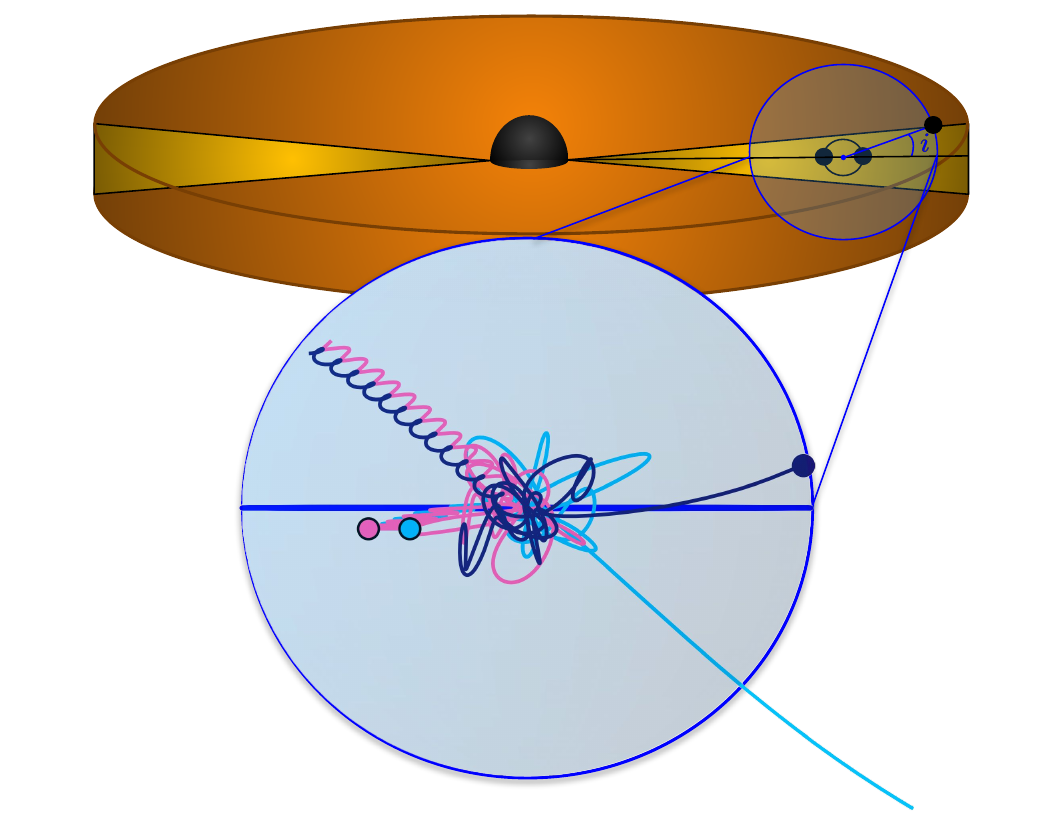}
\includegraphics[width=1.1\columnwidth]{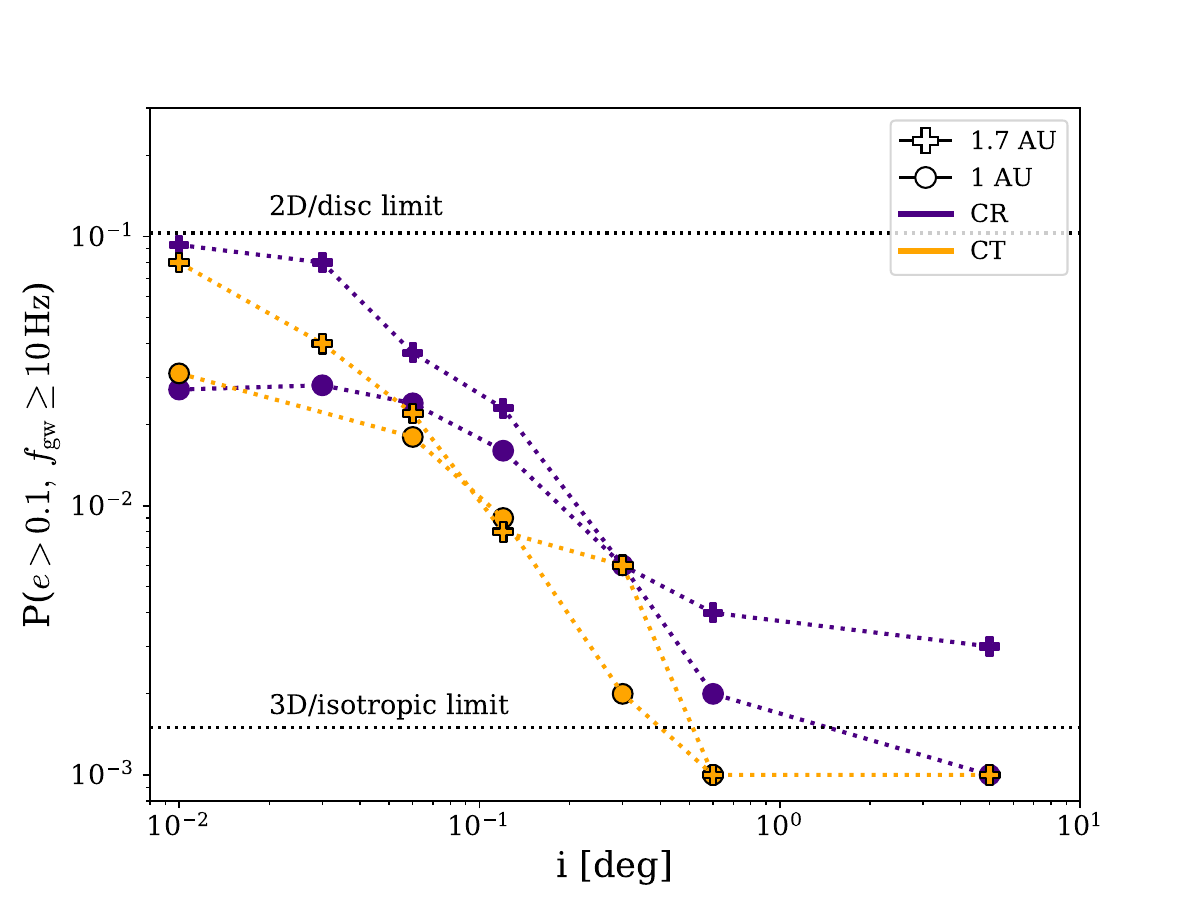}
\caption{
Top figure: Illustration of the setup we use for the out-of-plane interactions we explore in Sec. \ref{sec:3D}, which is similar to the setup shown in Fig. \ref{fig:setup}, but now also with inclination angle $i$. An example of an out-of-plane scattering is also shown in the large blue circle indicating the Hill sphere. 
We here initiate the single object at the Hill sphere at an angle $i$ above the plane with the same velocity vector that we used in the co-planar case, i.e. we
assume that the velocity perpendicular to the plane is equal to zero when the simulation starts. This setup is used to explore and quantify how the vertical oscillations and relative orbital fluctuations
there are in the disc impact the formation of eccentric mergers. 
Bottom figure: Probability of eccentric mergers as a function of $i$ for the out-of-plane scatterings in both CR (purple) and CT (orange) configurations for two SMAs (1.7 AU and 1 AU). The dashed lines represent the 2D (disc) and 3D (isotropic) limits for eccentric mergers. 
\label{fig:prob_vs_i}}
\end{figure}

\subsection{Out-of-plane interactions}\label{sec:3D}

Until now we have focused on properties and outcomes of the co-planar setup, where all four objects (SMBH, binary+single)
are interacting in the same plane, as this configuration gives rise to the most distinct outcomes relative to the well studied isotropic case \citep{Samsing14,Samsing18}.
However, there are different factors that can result in a non-fully 2-D distribution of BHs embedded in the disc. When BHs in the nuclear star cluster are captured by the disc due to drag force processes they do not necessarily get aligned in fully co-planar orbits \citep[e.g.][]{Fabj20,Nasim22,2024MNRAS.528.4958W}, dynamical heating will occur if a larger population of the disc is present as discussed in \citet{2017MNRAS.464..946S}, and 
the orbital evolution of objects inside the disc is believed to be subject to e.g. gas turbulence \citep{2024MNRAS.533.1766W}. All these processes ultimately result in mutual interactions that are not co-planar, which we refer to as out-of-plane
interactions. 

We quantify the out-of-plane interaction by the inclination angle, $i$, which refers to the angle between the incoming single relative to the orbital
BBH and disc plane. As shown in the top part of Fig. \ref{fig:prob_vs_i}, we place the single incoming BH at the Hill sphere at angle $i$ above the disc-plane, with a velocity vector that is equivalent to the co-planar setup,
i.e. we consider the situation where the single BH initially has zero velocity in the direction perpendicular to the disc. A similar 
setup was explored in \citetalias{Samsing22}, but in their case the single BH was assumed to be
coming from infinity in a plane with inclination $i$ relative to the BBH orbital plane, in which case the single BH
passes through the Hill sphere approximately with the same angle $i$, but with a slightly different angular momentum.
In \citetalias{Samsing22} it was shown that the outcome probability of eccentric mergers strongly depends on $i$, with nearly one order-of-magnitude
decrease when the angle has opened to $1^{\circ}$ relative to
the $0^{\circ}$ co-planar case. This strong dependence on the orbital plane inclination angle naturally questions how the results actually
depend on the assumption that the single BH approached from infinity as normally assumed in
scattering systems. It could be that if the object is initiated at the Hill sphere, and not propagated to the Hill sphere from far away, the dependence on $i$ is much weaker, as what really matters is how much angular momentum the single BH brings. With this motivation, we now consider the results shown in Fig. \ref{fig:prob_vs_i}, which shows the probability for
eccentric mergers ($e>0.1$ with a peak frequency $f_{\rm gw} \geq 10$ Hz), resulting from our
out-of-plane \textit{Setup I} where the single starts at the Hill sphere at an angle $i$ above the plane. In Fig. \ref{fig:prob_vs_i} the BH masses are equal and set to $15 \, M_{\odot}$. We show in total four different setups characterized by two different SMAs (1.7 AU, cross symbols and 1 AU, circles)
and for CR (purple) and CT (orange) binaries. As seen, the general trend is consistent with \citetalias{Samsing22}, i.e. at angle $10^{-2}$ the probabilities converge to the value found in the co-planar case shown in Fig. \ref{fig:prob}, with a rapid decrease by almost 2-3 orders of magnitude once $i$
passes $1^{\circ}-10^{\circ}$. Even in this setup, the probability for eccentric mergers therefore still seems to be very sensitive to the inclination angle, which adds further motivation for studying how the gas has an impact on the alignment of the objects in the disc. This is particularly important
when considering the binaries surviving the interaction, as they are scattered both out of the disc-plane
and have their orbital momentum tilted \citepalias{Samsing22}. Many of these binaries are likely to undergo subsequent interactions inside the disc, but how aligned they are in the disc at the time of interaction depends on the alignment timescales relative to the interaction timescale, as well as the properties of the binary after interaction. 
The subsequent question is naturally what the expected out-of-plane angle could be in AGN-disc like environments. Assuming the binary orbital angular moment is perpendicular to the disc angular momentum, a simple estimate can be made in the case
the Hill sphere is larger than the disc height,
in which case the maximum opening angle is around the AGN-disc height divided by the Hill radius, i.e. $max(i) \approx h/R_H \approx (h/R) \times \left(M/m\right)^{1/3}$. The maximum opening angle can span a range of different values, depending on both aspect-ratio and mass ratio.
For example,
for a system described by $m=10^{1} M_{\odot}$, $M=10^{7} M_{\odot}$, then $max(i) \approx h/R_{\rm H} \times 10^{2}$, 
i.e. for thin-disc models with $h / R \sim 10^{-3}$ like \citet{Shakura73}, the characteristic angle could very well be $<1 ^{\circ}$. Reading from the bottom panel in Fig. \ref{fig:prob_vs_i}, this would still mean that
eccentric mergers are likely to form in over-abundance in disc environments compared to isotropic environments. However, for more
realistic disc models, this simple estimate could vary throughout the disc, reaching values of $h/R \sim 0.1$ \citep[e.g.][]{SG,TQM}. This structure would make some regions more prone to forming eccentric mergers
than others.

\section{Conclusions and caveats} \label{sec:Discussion}

In this paper we have explored the outcomes from binary-single interactions between migrating BBHs and single BHs in AGN disc-like environments in the presence of an SMBH.
In particular, we have studied how the fraction of eccentric mergers depends on the BBH SMA relative to the Hill sphere created by the tidal field of the SMBH, the BBH direction of rotation
relative to the disc, as well as the relative inclination angle between the BBH and the single BH
as they approach each other through their Hill sphere. We have further presented the first analytical solution to how a tidal field in general terms has an impact on the evolution
of resonating binary-single scatterings, which are one of the main pathways
for producing eccentric mergers. In the following list we summarize our main conclusions and discuss the possible implications of adding the effect of a gaseous disc.

\begin{itemize}
    
    \item {\bf Theory of 3-body interactions in a tidal field.}
     When binary-single objects interact, they often undergo a large number of resonant states that are characterized by long
     excursions of the single relative to the binary. Each of these IMS states has been shown to be one of the main
     pathways to form eccentric mergers in systems where such interactions can take place. In Sec. \ref{sec:tid fields}, we have extended the theory
     of how such eccentric 3-body mergers form in resonant interactions, by including the effect from a tidal field characterized by
     a Hill sphere boundary. By the use of statistical theory for how the energy and angular momentum is distributed in chaotic encounters, based on works by \citet{2006tbp..book.....V, 2019Natur.576..406S},
     we showed that the probability for 3-body mergers, i.e. mergers forming during the interactions, are generally reduced by a factor of $(1-2 a_0 /R_{\rm H})^2$, where $a_0$ is the initial BBH semi-major axis 
     and $R_{\rm H}$ is the Hill sphere (see Eq. \ref{eq:Rhill}). While true chaotic interactions are difficult to achieve in environments with a strong nearby perturber,
     we identify this as the first indication of how a nearby SMBH might impact the fraction of eccentric mergers. Fig. \ref{fig:linefit}, shows this correction factor
     applied on the results from \citetalias{Samsing22}.
    
    \item {\bf PN-Simulations of eccentric mergers under the influence of the SMBH.}
    We performed a systematic study of PN scatterings between a BBH and a single incoming BH taking place near the SMBH. In this setup, we especially explored the
    formation probability of eccentric BBH mergers ($e$ > 0.1 at $f_{\rm gw}$ > 10 Hz), and how this depends on the SMA of the initial BBH $(a_0)$ relative to the effective Hill sphere of the
    SMBH ($R_{\rm H}$), its orbital angular momentum relative to its direction around the SMBH (co- and counter-rotating), the location of the single relative to the
    BBH (\textit{Setup I} and \textit{Setup II}, see Fig. \ref{fig:setup}), as well as the initial inclination between the BBH and single BH. In all scatterings we initiated the single BH at the Hill sphere, to
    emulate the likely process that takes place when a BBH and a single black hole meet in an AGN-disc environment through disc migration. \citetalias{Samsing22} showed that three-body scatterings in disc-like configurations with no SMBH included,
    produce a high fraction of eccentric mergers. With more realistic setups we here find that eccentric mergers are indeed also found to be produced
    in large numbers even when the interactions take place close to the SMBH. The reason is partly that many of the near co-planar interactions produce mergers that
    are formed through prompt interactions (see Fig. \ref{fig:zoom_1_int}), that are not greatly influenced by the SMBH upon formation.
    Finally, as shown in Fig. \ref{fig:fgw_dist}, many of the 3-body mergers do actually start with a GW peak frequency that can be far above 10 Hz, hinting that a significant fraction might show up as
    bursts or prompt mergers, rather than classical eccentric inspirals, where the system evolves from eccentric to circular over hundreds of cycles. This result is in agreement with the analysis of \citet{Trani23} and \citet{2023arXiv231003801R}, where for the latter they explore the effects of tidal forces coming from the SMBH on binary capture in AGNs. These findings further motivate
    searches for burst-like GW signals.
    
    \item {\bf Merger types and phase-space distributions.}
    We studied the distribution of outcomes as a function of SMA, and orbital phase angle $\phi_0$ for different setups,
    and found a clear structure that propagates to our main results of the probability for eccentric mergers as shown in Fig. \ref{fig:zoom_2_prob}. Generally, we found that
    when the SMA of the BBH is much smaller than the Hill sphere, the scatterings nearly lose information about the ICs, and the results therefore approach the one
    found by \citetalias{Samsing22} that by construction explored outcomes from a much less constrained setup (the single coming in isotropically in a plane, where we e.g. start always at the same
    point at the Hill sphere). On the other hand, when the SMA increases and starts to approach the size of $R_{\rm H}$, the outcomes strongly depend on the exact angular phase relative to the SMA,
    with an approximate periodicity arising from considering integer values of the interaction time relative to the BBH orbital time (see Eq. \ref{eq:resma} and Fig. \ref{fig:prob}). Despite these strong
    dependencies when approaching the Hill sphere, we find that the probability for eccentric mergers in this region as well approximately matches the one found by \citetalias{Samsing22}, but with significant
    fluctuations coming from the underlying phase-space structure. In addition, both CT and CR for \textit{Setup I}  and \textit{Setup II} all showed similar outcome distribution.

    \item {\bf Importance of orbital inclinations.}
     No interactions are expected to be perfectly co-planar, studying how our results depend on the initial inclination angle between the BBH and single when the single enters the Hill sphere (angle $i$
     as seen in Fig. \ref{fig:prob_vs_i}), is therefore of major importance. In Sec. \ref{sec:3D}, we explored the fraction of eccentric BBH mergers as a function of the initial inclination angle $i$, and found
     that across the different constrained setups, the probability does falls off quickly with $i$ as was also pointed out in \citetalias{Samsing22}. All AGN-disc environments are different from each other,
     but we argue that for thin disc models, in regions where the maximum opening angle is $i \sim 1^{\circ}$, a relatively high fraction of eccentric mergers are likely
     to be produced.

     \item {\bf Potential impact of gas in the accretion disc.}
     As discussed, we do not include gaseous effects in our analysis but acknowledge their potential impact on the presented results and on the overall fraction of eccentric mergers. A relevant question to be explored is if the additional gas drag force term keeps the objects contained in the Hill sphere, ultimately leading to more chaotic interactions in the co-rotating and counter-rotating case as a contrast to the high fraction of prompt interaction regions we observe from e.g. Fig. \ref{fig:phaseplot} \& \ref{fig:zoom_1_int}.
     In addition, a full dynamical model of migrating BHs inside AGN-disc-like environments is necessary to state what relevant values of $R_{\rm H}$ and SMA are to be expected, and therefore how many interactions can be considered chaotic or not.
     As far as the nature and outlook of mergers is concerned, the scattering seen in the bottom panel of Fig. \ref{fig:combo} might in comparison not be much affected by gas, as the binary undergoing merger is formed promptly. Therefore, we speculate that gas might not have a great impact on the formation of such prompt mergers, which could hint that our results might also hold when gaseous effects are included.
     There are however, several reasons to believe that our considerations and assumptions from above need to be adjusted, as the relevant setup that we consider involves binaries that encounter singles as a result of migration through an AGN-disc \citep[e.g.][]{2019ApJ...878...85S,2020ApJ...903..133S,Tagawa20}. Lastly, whether gas friction is expected to damp or pump the eccentricity of the binary is a non-trivial answer, as its evolution depends on a number of factors of the parameter space, such as mass ratio or whether binary is in prograde or retrograde motion with respect to the gas \citep[e.g.][]{Li22a, 2023MNRAS.522.1881L, Li22c,2023arXiv231003832D, 2024arXiv240117355V}.

\end{itemize}

\section*{Acknowledgements} \label{sec:Acknowledgements}
The authors thank the anonymous referee for the useful suggestions and comments.
GF thanks the GW-Astro group at NBIA, in particular Martin Pessah and Daniel J. D'Orazio, for insightful discussions and constant support, as well as the TIDY-NYC group for interesting discussions. The Tycho supercomputer hosted at the SCIENCE HPC center at the University of Copenhagen was used for performing the simulations presented in the paper. This work was funded by the Villum
Fonden Grant No. 29466 and ERC Starting Grant no. 101043143.

\section*{Data Availability}
The data underlying this article will be shared on request to the corresponding author.

\bibliographystyle{mnras}
\bibliography{references}

\begin{thebibliography}{}
\makeatletter
\relax
\def\mn@urlcharsother{\let\do\@makeother \do\$\do\&\do\#\do\^\do\_\do\%\do\~}
\def\mn@doi{\begingroup\mn@urlcharsother \@ifnextchar [ {\mn@doi@}
  {\mn@doi@[]}}
\def\mn@doi@[#1]#2{\def\@tempa{#1}\ifx\@tempa\@empty \href
  {http://dx.doi.org/#2} {doi:#2}\else \href {http://dx.doi.org/#2} {#1}\fi
  \endgroup}
\def\mn@eprint#1#2{\mn@eprint@#1:#2::\@nil}
\def\mn@eprint@arXiv#1{\href {http://arxiv.org/abs/#1} {{\tt arXiv:#1}}}
\def\mn@eprint@dblp#1{\href {http://dblp.uni-trier.de/rec/bibtex/#1.xml}
  {dblp:#1}}
\def\mn@eprint@#1:#2:#3:#4\@nil{\def\@tempa {#1}\def\@tempb {#2}\def\@tempc
  {#3}\ifx \@tempc \@empty \let \@tempc \@tempb \let \@tempb \@tempa \fi \ifx
  \@tempb \@empty \def\@tempb {arXiv}\fi \@ifundefined
  {mn@eprint@\@tempb}{\@tempb:\@tempc}{\expandafter \expandafter \csname
  mn@eprint@\@tempb\endcsname \expandafter{\@tempc}}}

\bibitem[\protect\citeauthoryear{{Abbott} et~al.,}{{Abbott}
  et~al.}{2016a}]{2016PhRvX...6d1015A}
{Abbott} B.~P.,  et~al., 2016a, \mn@doi [Physical Review X]
  {10.1103/PhysRevX.6.041015}, \href
  {http://adsabs.harvard.edu/abs/2016PhRvX...6d1015A} {6, 041015}

\bibitem[\protect\citeauthoryear{{Abbott} et~al.,}{{Abbott}
  et~al.}{2016b}]{2016PhRvL.116f1102A}
{Abbott} B.~P.,  et~al., 2016b, \mn@doi [Physical Review Letters]
  {10.1103/PhysRevLett.116.061102}, \href
  {http://adsabs.harvard.edu/abs/2016PhRvL.116f1102A} {116, 061102}

\bibitem[\protect\citeauthoryear{{Abbott} et~al.,}{{Abbott}
  et~al.}{2016c}]{2016PhRvL.116x1103A}
{Abbott} B.~P.,  et~al., 2016c, \mn@doi [Physical Review Letters]
  {10.1103/PhysRevLett.116.241103}, \href
  {http://adsabs.harvard.edu/abs/2016PhRvL.116x1103A} {116, 241103}

\bibitem[\protect\citeauthoryear{{Abbott} et~al.,}{{Abbott}
  et~al.}{2017a}]{2017PhRvL.118v1101A}
{Abbott} B.~P.,  et~al., 2017a, \mn@doi [Physical Review Letters]
  {10.1103/PhysRevLett.118.221101}, \href
  {http://adsabs.harvard.edu/abs/2017PhRvL.118v1101A} {118, 221101}

\bibitem[\protect\citeauthoryear{{Abbott} et~al.,}{{Abbott}
  et~al.}{2017b}]{2017PhRvL.119n1101A}
{Abbott} B.~P.,  et~al., 2017b, \mn@doi [Physical Review Letters]
  {10.1103/PhysRevLett.119.141101}, \href
  {http://adsabs.harvard.edu/abs/2017PhRvL.119n1101A} {119, 141101}

\bibitem[\protect\citeauthoryear{{Abbott} et~al.,}{{Abbott}
  et~al.}{2017c}]{2017PhRvL.119p1101A}
{Abbott} B.~P.,  et~al., 2017c, \mn@doi [Physical Review Letters]
  {10.1103/PhysRevLett.119.161101}, \href
  {http://adsabs.harvard.edu/abs/2017PhRvL.119p1101A} {119, 161101}

\bibitem[\protect\citeauthoryear{{Abbott} et~al.,}{{Abbott}
  et~al.}{2020a}]{2020CQGra..37e5002A}
{Abbott} B.~P.,  et~al., 2020a, \mn@doi [Classical and Quantum Gravity]
  {10.1088/1361-6382/ab685e}, \href
  {https://ui.adsabs.harvard.edu/abs/2020CQGra..37e5002A} {37, 055002}

\bibitem[\protect\citeauthoryear{{Abbott} et~al.,}{{Abbott}
  et~al.}{2020b}]{GW19a}
{Abbott} R.,  et~al., 2020b, \mn@doi [\prl] {10.1103/PhysRevLett.125.101102},
  \href {https://ui.adsabs.harvard.edu/abs/2020PhRvL.125j1102A} {125, 101102}

\bibitem[\protect\citeauthoryear{{Abbott} et~al.,}{{Abbott}
  et~al.}{2020c}]{GW19b}
{Abbott} R.,  et~al., 2020c, \mn@doi [\apjl] {10.3847/2041-8213/aba493}, \href
  {https://ui.adsabs.harvard.edu/abs/2020ApJ...900L..13A} {900, L13}

\bibitem[\protect\citeauthoryear{{Abbott} et~al.,}{{Abbott}
  et~al.}{2023}]{2023PhRvX..13d1039A}
{Abbott} R.,  et~al., 2023, \mn@doi [Physical Review X]
  {10.1103/PhysRevX.13.041039}, \href
  {https://ui.adsabs.harvard.edu/abs/2023PhRvX..13d1039A} {13, 041039}

\bibitem[\protect\citeauthoryear{{Antonini} \& {Rasio}}{{Antonini} \&
  {Rasio}}{2016}]{2016ApJ...831..187A}
{Antonini} F.,  {Rasio} F.~A.,  2016, \mn@doi [\apj]
  {10.3847/0004-637X/831/2/187}, \href
  {http://adsabs.harvard.edu/abs/2016ApJ...831..187A} {831, 187}

\bibitem[\protect\citeauthoryear{{Artale}, {Mapelli}, {Giacobbo}, {Sabha},
  {Spera}, {Santoliquido}  \& {Bressan}}{{Artale}
  et~al.}{2019}]{2019MNRAS.487.1675A}
{Artale} M.~C.,  {Mapelli} M.,  {Giacobbo} N.,  {Sabha} N.~B.,  {Spera} M.,
  {Santoliquido} F.,   {Bressan} A.,  2019, \mn@doi [\mnras]
  {10.1093/mnras/stz1382}, \href
  {https://ui.adsabs.harvard.edu/abs/2019MNRAS.487.1675A} {487, 1675}

\bibitem[\protect\citeauthoryear{{Askar}, {Szkudlarek}, {Gondek-Rosi{\'n}ska},
  {Giersz}  \& {Bulik}}{{Askar} et~al.}{2017}]{2017MNRAS.464L..36A}
{Askar} A.,  {Szkudlarek} M.,  {Gondek-Rosi{\'n}ska} D.,  {Giersz} M.,
  {Bulik} T.,  2017, \mn@doi [\mnras] {10.1093/mnrasl/slw177}, \href
  {http://adsabs.harvard.edu/abs/2017MNRAS.464L..36A} {464, L36}

\bibitem[\protect\citeauthoryear{{Bae}, {Kim}  \& {Lee}}{{Bae}
  et~al.}{2014}]{2014MNRAS.440.2714B}
{Bae} Y.-B.,  {Kim} C.,   {Lee} H.~M.,  2014, \mn@doi [\mnras]
  {10.1093/mnras/stu381}, \href
  {http://adsabs.harvard.edu/abs/2014MNRAS.440.2714B} {440, 2714}

\bibitem[\protect\citeauthoryear{{Banerjee}, {Baumgardt}  \&
  {Kroupa}}{{Banerjee} et~al.}{2010}]{2010MNRAS.402..371B}
{Banerjee} S.,  {Baumgardt} H.,   {Kroupa} P.,  2010, \mn@doi [\mnras]
  {10.1111/j.1365-2966.2009.15880.x}, \href
  {http://adsabs.harvard.edu/abs/2010MNRAS.402..371B} {402, 371}

\bibitem[\protect\citeauthoryear{{Bartos}, {Kocsis}, {Haiman}  \&
  {M{\'a}rka}}{{Bartos} et~al.}{2017}]{2017ApJ...835..165B}
{Bartos} I.,  {Kocsis} B.,  {Haiman} Z.,   {M{\'a}rka} S.,  2017, \mn@doi
  [\apj] {10.3847/1538-4357/835/2/165}, \href
  {http://adsabs.harvard.edu/abs/2017ApJ...835..165B} {835, 165}

\bibitem[\protect\citeauthoryear{{Belczynski}, {Holz}, {Bulik}  \&
  {O'Shaughnessy}}{{Belczynski} et~al.}{2016a}]{2016Natur.534..512B}
{Belczynski} K.,  {Holz} D.~E.,  {Bulik} T.,   {O'Shaughnessy} R.,  2016a,
  \mn@doi [\nat] {10.1038/nature18322}, \href
  {http://adsabs.harvard.edu/abs/2016Natur.534..512B} {534, 512}

\bibitem[\protect\citeauthoryear{{Belczynski}, {Repetto}, {Holz},
  {O'Shaughnessy}, {Bulik}, {Berti}, {Fryer}  \& {Dominik}}{{Belczynski}
  et~al.}{2016b}]{2016ApJ...819..108B}
{Belczynski} K.,  {Repetto} S.,  {Holz} D.~E.,  {O'Shaughnessy} R.,  {Bulik}
  T.,  {Berti} E.,  {Fryer} C.,   {Dominik} M.,  2016b, \mn@doi [\apj]
  {10.3847/0004-637X/819/2/108}, \href
  {http://adsabs.harvard.edu/abs/2016ApJ...819..108B} {819, 108}

\bibitem[\protect\citeauthoryear{{Bird}, {Cholis}, {Mu{\~n}oz},
  {Ali-Ha{\"i}moud}, {Kamionkowski}, {Kovetz}, {Raccanelli}  \& {Riess}}{{Bird}
  et~al.}{2016}]{2016PhRvL.116t1301B}
{Bird} S.,  {Cholis} I.,  {Mu{\~n}oz} J.~B.,  {Ali-Ha{\"i}moud} Y.,
  {Kamionkowski} M.,  {Kovetz} E.~D.,  {Raccanelli} A.,   {Riess} A.~G.,  2016,
  \mn@doi [Physical Review Letters] {10.1103/PhysRevLett.116.201301}, \href
  {http://adsabs.harvard.edu/abs/2016PhRvL.116t1301B} {116, 201301}

\bibitem[\protect\citeauthoryear{{Blanchet}}{{Blanchet}}{2006}]{Blanchet06}
{Blanchet} L.,  2006, \mn@doi [Living Reviews in Relativity]
  {10.12942/lrr-2006-4}, \href
  {https://ui.adsabs.harvard.edu/abs/2006LRR.....9....4B} {9, 4}

\bibitem[\protect\citeauthoryear{{Blanchet}}{{Blanchet}}{2014}]{Blanchet14}
{Blanchet} L.,  2014, \mn@doi [Living Reviews in Relativity]
  {10.12942/lrr-2014-2}, \href
  {https://ui.adsabs.harvard.edu/abs/2014LRR....17....2B} {17, 2}

\bibitem[\protect\citeauthoryear{{Boekholt}, {Rowan}  \& {Kocsis}}{{Boekholt}
  et~al.}{2023}]{Boekholt23}
{Boekholt} T. C.~N.,  {Rowan} C.,   {Kocsis} B.,  2023, \mn@doi [\mnras]
  {10.1093/mnras/stac3495}, \href
  {https://ui.adsabs.harvard.edu/abs/2023MNRAS.518.5653B} {518, 5653}

\bibitem[\protect\citeauthoryear{{Calcino}, {Dempsey}, {Dittmann}  \&
  {Li}}{{Calcino} et~al.}{2024}]{2024ApJ...970..107C}
{Calcino} J.,  {Dempsey} A.~M.,  {Dittmann} A.~J.,   {Li} H.,  2024, \mn@doi
  [\apj] {10.3847/1538-4357/ad4a53}, \href
  {https://ui.adsabs.harvard.edu/abs/2024ApJ...970..107C} {970, 107}

\bibitem[\protect\citeauthoryear{{Cantiello}, {Jermyn}  \& {Lin}}{{Cantiello}
  et~al.}{2021}]{Cantiello21}
{Cantiello} M.,  {Jermyn} A.~S.,   {Lin} D. N.~C.,  2021, \mn@doi [\apj]
  {10.3847/1538-4357/abdf4f}, \href
  {https://ui.adsabs.harvard.edu/abs/2021ApJ...910...94C} {910, 94}

\bibitem[\protect\citeauthoryear{{Carr}, {K{\"u}hnel}  \& {Sandstad}}{{Carr}
  et~al.}{2016}]{2016PhRvD..94h3504C}
{Carr} B.,  {K{\"u}hnel} F.,   {Sandstad} M.,  2016, \mn@doi [\prd]
  {10.1103/PhysRevD.94.083504}, \href
  {http://adsabs.harvard.edu/abs/2016PhRvD..94h3504C} {94, 083504}

\bibitem[\protect\citeauthoryear{{Chen} \& {Amaro-Seoane}}{{Chen} \&
  {Amaro-Seoane}}{2017}]{2017ApJ...842L...2C}
{Chen} X.,  {Amaro-Seoane} P.,  2017, \mn@doi [\apjl]
  {10.3847/2041-8213/aa74ce}, \href
  {http://adsabs.harvard.edu/abs/2017ApJ...842L...2C} {842, L2}

\bibitem[\protect\citeauthoryear{{Cholis}, {Kovetz}, {Ali-Ha{\"i}moud}, {Bird},
  {Kamionkowski}, {Mu{\~n}oz}  \& {Raccanelli}}{{Cholis}
  et~al.}{2016}]{2016PhRvD..94h4013C}
{Cholis} I.,  {Kovetz} E.~D.,  {Ali-Ha{\"i}moud} Y.,  {Bird} S.,
  {Kamionkowski} M.,  {Mu{\~n}oz} J.~B.,   {Raccanelli} A.,  2016, \mn@doi
  [\prd] {10.1103/PhysRevD.94.084013}, \href
  {http://adsabs.harvard.edu/abs/2016PhRvD..94h4013C} {94, 084013}

\bibitem[\protect\citeauthoryear{{D'Orazio} \& {Loeb}}{{D'Orazio} \&
  {Loeb}}{2018}]{DOrazioLoeb:2017}
{D'Orazio} D.~J.,  {Loeb} A.,  2018, \mn@doi [\prd]
  {10.1103/PhysRevD.97.083008}, \href
  {https://ui.adsabs.harvard.edu/abs/2018PhRvD..97h3008D} {97, 083008}

\bibitem[\protect\citeauthoryear{{D'Orazio} \& {Loeb}}{{D'Orazio} \&
  {Loeb}}{2020}]{2020PhRvD.101h3031D}
{D'Orazio} D.~J.,  {Loeb} A.,  2020, \mn@doi [\prd]
  {10.1103/PhysRevD.101.083031}, \href
  {https://ui.adsabs.harvard.edu/abs/2020PhRvD.101h3031D} {101, 083031}

\bibitem[\protect\citeauthoryear{{D'Orazio} \& {Samsing}}{{D'Orazio} \&
  {Samsing}}{2018a}]{2018MNRAS.481.4775D}
{D'Orazio} D.~J.,  {Samsing} J.,  2018a, \mn@doi [\mnras]
  {10.1093/mnras/sty2568}, \href
  {http://adsabs.harvard.edu/abs/2018MNRAS.481.4775D} {481, 4775}

\bibitem[\protect\citeauthoryear{{D'Orazio} \& {Samsing}}{{D'Orazio} \&
  {Samsing}}{2018b}]{DJD18}
{D'Orazio} D.~J.,  {Samsing} J.,  2018b, \mn@doi [\mnras]
  {10.1093/mnras/sty2568}, \href
  {https://ui.adsabs.harvard.edu/abs/2018MNRAS.481.4775D} {481, 4775}

\bibitem[\protect\citeauthoryear{{DeLaurentiis}, {Epstein-Martin}  \&
  {Haiman}}{{DeLaurentiis} et~al.}{2023}]{Delaurentiis23}
{DeLaurentiis} S.,  {Epstein-Martin} M.,   {Haiman} Z.,  2023, \mn@doi [\mnras]
  {10.1093/mnras/stad1412}, \href
  {https://ui.adsabs.harvard.edu/abs/2023MNRAS.523.1126D} {523, 1126}

\bibitem[\protect\citeauthoryear{{Dempsey}, {Li}, {Mishra}  \& {Li}}{{Dempsey}
  et~al.}{2022}]{2022ApJ...940..155D}
{Dempsey} A.~M.,  {Li} H.,  {Mishra} B.,   {Li} S.,  2022, \mn@doi [\apj]
  {10.3847/1538-4357/ac9d92}, \href
  {https://ui.adsabs.harvard.edu/abs/2022ApJ...940..155D} {940, 155}

\bibitem[\protect\citeauthoryear{{Dittmann}, {Cantiello}  \&
  {Jermyn}}{{Dittmann} et~al.}{2021}]{2021ApJ...916...48D}
{Dittmann} A.~J.,  {Cantiello} M.,   {Jermyn} A.~S.,  2021, \mn@doi [\apj]
  {10.3847/1538-4357/ac042c}, \href
  {https://ui.adsabs.harvard.edu/abs/2021ApJ...916...48D} {916, 48}

\bibitem[\protect\citeauthoryear{{Dittmann}, {Dempsey}  \& {Li}}{{Dittmann}
  et~al.}{2023a}]{2023arXiv231003832D}
{Dittmann} A.~J.,  {Dempsey} A.~M.,   {Li} H.,  2023a, \mn@doi [arXiv e-prints]
  {10.48550/arXiv.2310.03832}, \href
  {https://ui.adsabs.harvard.edu/abs/2023arXiv231003832D} {p. arXiv:2310.03832}

\bibitem[\protect\citeauthoryear{{Dittmann}, {Jermyn}  \&
  {Cantiello}}{{Dittmann} et~al.}{2023b}]{2023ApJ...946...56D}
{Dittmann} A.~J.,  {Jermyn} A.~S.,   {Cantiello} M.,  2023b, \mn@doi [\apj]
  {10.3847/1538-4357/acacf2}, \href
  {https://ui.adsabs.harvard.edu/abs/2023ApJ...946...56D} {946, 56}

\bibitem[\protect\citeauthoryear{{Dittmann}, {Dempsey}  \& {Li}}{{Dittmann}
  et~al.}{2024}]{2024ApJ...964...61D}
{Dittmann} A.~J.,  {Dempsey} A.~M.,   {Li} H.,  2024, \mn@doi [\apj]
  {10.3847/1538-4357/ad23ce}, \href
  {https://ui.adsabs.harvard.edu/abs/2024ApJ...964...61D} {964, 61}

\bibitem[\protect\citeauthoryear{{Dominik}, {Belczynski}, {Fryer}, {Holz},
  {Berti}, {Bulik}, {Mandel}  \& {O'Shaughnessy}}{{Dominik}
  et~al.}{2012}]{2012ApJ...759...52D}
{Dominik} M.,  {Belczynski} K.,  {Fryer} C.,  {Holz} D.~E.,  {Berti} E.,
  {Bulik} T.,  {Mandel} I.,   {O'Shaughnessy} R.,  2012, \mn@doi [\apj]
  {10.1088/0004-637X/759/1/52}, \href
  {http://adsabs.harvard.edu/abs/2012ApJ...759...52D} {759, 52}

\bibitem[\protect\citeauthoryear{{Dominik}, {Belczynski}, {Fryer}, {Holz},
  {Berti}, {Bulik}, {Mandel}  \& {O'Shaughnessy}}{{Dominik}
  et~al.}{2013}]{2013ApJ...779...72D}
{Dominik} M.,  {Belczynski} K.,  {Fryer} C.,  {Holz} D.~E.,  {Berti} E.,
  {Bulik} T.,  {Mandel} I.,   {O'Shaughnessy} R.,  2013, \mn@doi [\apj]
  {10.1088/0004-637X/779/1/72}, \href
  {http://adsabs.harvard.edu/abs/2013ApJ...779...72D} {779, 72}

\bibitem[\protect\citeauthoryear{{Dominik} et~al.,}{{Dominik}
  et~al.}{2015}]{2015ApJ...806..263D}
{Dominik} M.,  et~al., 2015, \mn@doi [\apj] {10.1088/0004-637X/806/2/263},
  \href {http://adsabs.harvard.edu/abs/2015ApJ...806..263D} {806, 263}

\bibitem[\protect\citeauthoryear{{Fabj}, {Nasim}, {Caban}, {Ford}, {McKernan}
  \& {Bellovary}}{{Fabj} et~al.}{2020}]{Fabj20}
{Fabj} G.,  {Nasim} S.~S.,  {Caban} F.,  {Ford} K.~E.~S.,  {McKernan} B.,
  {Bellovary} J.~M.,  2020, \mn@doi [MNRAS] {10.1093/mnras/staa3004}, \href
  {https://ui.adsabs.harvard.edu/abs/2020arXiv200611229F} {499, 2608–2616}

\bibitem[\protect\citeauthoryear{{Fishbach} \& {Holz}}{{Fishbach} \&
  {Holz}}{2020}]{2020ApJ...904L..26F}
{Fishbach} M.,  {Holz} D.~E.,  2020, \mn@doi [\apjl]
  {10.3847/2041-8213/abc827}, \href
  {https://ui.adsabs.harvard.edu/abs/2020ApJ...904L..26F} {904, L26}

\bibitem[\protect\citeauthoryear{{Fishbach}, {Farr}  \& {Holz}}{{Fishbach}
  et~al.}{2020}]{2020ApJ...891L..31F}
{Fishbach} M.,  {Farr} W.~M.,   {Holz} D.~E.,  2020, \mn@doi [\apjl]
  {10.3847/2041-8213/ab77c9}, \href
  {https://ui.adsabs.harvard.edu/abs/2020ApJ...891L..31F} {891, L31}

\bibitem[\protect\citeauthoryear{{Gayathri} et~al.,}{{Gayathri}
  et~al.}{2022}]{2022NatAs...6..344G}
{Gayathri} V.,  et~al., 2022, \mn@doi [Nature Astronomy]
  {10.1038/s41550-021-01568-w}, \href
  {https://ui.adsabs.harvard.edu/abs/2022NatAs...6..344G} {6, 344}

\bibitem[\protect\citeauthoryear{{Generozov} \& {Perets}}{{Generozov} \&
  {Perets}}{2023}]{Generozov22}
{Generozov} A.,  {Perets} H.~B.,  2023, \mn@doi [\mnras]
  {10.1093/mnras/stad1016}, \href
  {https://ui.adsabs.harvard.edu/abs/2023MNRAS.522.1763G} {522, 1763}

\bibitem[\protect\citeauthoryear{{Gilbaum} \& {Stone}}{{Gilbaum} \&
  {Stone}}{2022}]{2022ApJ...928..191G}
{Gilbaum} S.,  {Stone} N.~C.,  2022, \mn@doi [\apj] {10.3847/1538-4357/ac4ded},
  \href {https://ui.adsabs.harvard.edu/abs/2022ApJ...928..191G} {928, 191}

\bibitem[\protect\citeauthoryear{{Ginat} \& {Perets}}{{Ginat} \&
  {Perets}}{2021}]{Ginat21}
{Ginat} Y.~B.,  {Perets} H.~B.,  2021, \mn@doi [\mnras]
  {10.1093/mnras/stab2565}, \href
  {https://ui.adsabs.harvard.edu/abs/2021MNRAS.508..190G} {508, 190}

\bibitem[\protect\citeauthoryear{{Graham} et~al.,}{{Graham}
  et~al.}{2020}]{Graham20}
{Graham} M.~J.,  et~al., 2020, \mn@doi [\prl] {10.1103/PhysRevLett.124.251102},
  \href {https://ui.adsabs.harvard.edu/abs/2020PhRvL.124y1102G} {124, 251102}

\bibitem[\protect\citeauthoryear{{Graham} et~al.,}{{Graham}
  et~al.}{2023}]{Graham23}
{Graham} M.~J.,  et~al., 2023, \mn@doi [\apj] {10.3847/1538-4357/aca480}, \href
  {https://ui.adsabs.harvard.edu/abs/2023ApJ...942...99G} {942, 99}

\bibitem[\protect\citeauthoryear{G{\"u}ltekin, Miller  \&
  Hamilton}{G{\"u}ltekin et~al.}{2006}]{2006ApJ...640..156G}
G{\"u}ltekin K.,  Miller M.~C.,   Hamilton D.~P.,  2006, \apj, 640, 156

\bibitem[\protect\citeauthoryear{{Hamers}, {Bar-Or}, {Petrovich}  \&
  {Antonini}}{{Hamers} et~al.}{2018}]{2018ApJ...865....2H}
{Hamers} A.~S.,  {Bar-Or} B.,  {Petrovich} C.,   {Antonini} F.,  2018, \mn@doi
  [\apj] {10.3847/1538-4357/aadae2}, \href
  {https://ui.adsabs.harvard.edu/abs/2018ApJ...865....2H} {865, 2}

\bibitem[\protect\citeauthoryear{{Hamers}, {Fragione}, {Neunteufel}  \&
  {Kocsis}}{{Hamers} et~al.}{2021}]{2021MNRAS.506.5345H}
{Hamers} A.~S.,  {Fragione} G.,  {Neunteufel} P.,   {Kocsis} B.,  2021, \mn@doi
  [\mnras] {10.1093/mnras/stab2136}, \href
  {https://ui.adsabs.harvard.edu/abs/2021MNRAS.506.5345H} {506, 5345}

\bibitem[\protect\citeauthoryear{Heggie}{Heggie}{1975}]{Heggie:1975uy}
Heggie D.~C.,  1975, \mnras, 173, 729

\bibitem[\protect\citeauthoryear{{Hendriks}, {Zwick}  \& {Samsing}}{{Hendriks}
  et~al.}{2024}]{2024arXiv240804603H}
{Hendriks} K.,  {Zwick} L.,   {Samsing} J.,  2024, \mn@doi [arXiv e-prints]
  {10.48550/arXiv.2408.04603}, \href
  {https://ui.adsabs.harvard.edu/abs/2024arXiv240804603H} {p. arXiv:2408.04603}

\bibitem[\protect\citeauthoryear{{Hoang}, {Naoz}, {Kocsis}, {Rasio}  \&
  {Dosopoulou}}{{Hoang} et~al.}{2018}]{2017arXiv170609896H}
{Hoang} B.-M.,  {Naoz} S.,  {Kocsis} B.,  {Rasio} F.~A.,   {Dosopoulou} F.,
  2018, \mn@doi [\apj] {10.3847/1538-4357/aaafce}, \href
  {https://ui.adsabs.harvard.edu/abs/2018ApJ...856..140H} {856, 140}

\bibitem[\protect\citeauthoryear{{Hong} \& {Lee}}{{Hong} \&
  {Lee}}{2015}]{2015MNRAS.448..754H}
{Hong} J.,  {Lee} H.~M.,  2015, \mn@doi [\mnras] {10.1093/mnras/stv035}, \href
  {http://adsabs.harvard.edu/abs/2015MNRAS.448..754H} {448, 754}

\bibitem[\protect\citeauthoryear{{Hotokezaka} \& {Piran}}{{Hotokezaka} \&
  {Piran}}{2017}]{Hotokezaka2017ApJ}
{Hotokezaka} K.,  {Piran} T.,  2017, \mn@doi [\apj] {10.3847/1538-4357/aa6f61},
  \href {https://ui.adsabs.harvard.edu/abs/2017ApJ...842..111H} {842, 111}

\bibitem[\protect\citeauthoryear{{Inayoshi}, {Tamanini}, {Caprini}  \&
  {Haiman}}{{Inayoshi} et~al.}{2017}]{2017PhRvD..96f3014I}
{Inayoshi} K.,  {Tamanini} N.,  {Caprini} C.,   {Haiman} Z.,  2017, \mn@doi
  [\prd] {10.1103/PhysRevD.96.063014}, \href
  {https://ui.adsabs.harvard.edu/abs/2017PhRvD..96f3014I} {96, 063014}

\bibitem[\protect\citeauthoryear{{Janiuk}, {Bejger}, {Charzy{\'n}ski}  \&
  {Sukova}}{{Janiuk} et~al.}{2017}]{Janiuk+2017}
{Janiuk} A.,  {Bejger} M.,  {Charzy{\'n}ski} S.,   {Sukova} P.,  2017, \mn@doi
  [\na] {10.1016/j.newast.2016.08.002}, \href
  {https://ui.adsabs.harvard.edu/abs/2017NewA...51....7J} {51, 7}

\bibitem[\protect\citeauthoryear{{Jermyn}, {Dittmann}, {Cantiello}  \&
  {Perna}}{{Jermyn} et~al.}{2021}]{Jermyn21}
{Jermyn} A.~S.,  {Dittmann} A.~J.,  {Cantiello} M.,   {Perna} R.,  2021,
  \mn@doi [\apj] {10.3847/1538-4357/abfb67}, \href
  {https://ui.adsabs.harvard.edu/abs/2021ApJ...914..105J} {914, 105}

\bibitem[\protect\citeauthoryear{{Jermyn}, {Dittmann}, {McKernan}, {Ford}  \&
  {Cantiello}}{{Jermyn} et~al.}{2022}]{2022ApJ...929..133J}
{Jermyn} A.~S.,  {Dittmann} A.~J.,  {McKernan} B.,  {Ford} K.~E.~S.,
  {Cantiello} M.,  2022, \mn@doi [\apj] {10.3847/1538-4357/ac5d40}, \href
  {https://ui.adsabs.harvard.edu/abs/2022ApJ...929..133J} {929, 133}

\bibitem[\protect\citeauthoryear{{Kalogera}}{{Kalogera}}{2000}]{2000ApJ...541..319K}
{Kalogera} V.,  2000, \mn@doi [\apj] {10.1086/309400}, \href
  {http://adsabs.harvard.edu/abs/2000ApJ...541..319K} {541, 319}

\bibitem[\protect\citeauthoryear{{Kinugawa}, {Inayoshi}, {Hotokezaka},
  {Nakauchi}  \& {Nakamura}}{{Kinugawa} et~al.}{2014}]{Kinugawa2014MNRAS}
{Kinugawa} T.,  {Inayoshi} K.,  {Hotokezaka} K.,  {Nakauchi} D.,   {Nakamura}
  T.,  2014, \mn@doi [\mnras] {10.1093/mnras/stu1022}, \href
  {https://ui.adsabs.harvard.edu/abs/2014MNRAS.442.2963K} {442, 2963}

\bibitem[\protect\citeauthoryear{{Klimenko} et~al.,}{{Klimenko}
  et~al.}{2016}]{2016PhRvD..93d2004K}
{Klimenko} S.,  et~al., 2016, \mn@doi [\prd] {10.1103/PhysRevD.93.042004},
  \href {https://ui.adsabs.harvard.edu/abs/2016PhRvD..93d2004K} {93, 042004}

\bibitem[\protect\citeauthoryear{{Kremer} et~al.,}{{Kremer}
  et~al.}{2019a}]{2019PhRvD..99f3003K}
{Kremer} K.,  et~al., 2019a, \mn@doi [\prd] {10.1103/PhysRevD.99.063003}, \href
  {https://ui.adsabs.harvard.edu/abs/2019PhRvD..99f3003K} {99, 063003}

\bibitem[\protect\citeauthoryear{{Kremer}, {Lu}, {Rodriguez}, {Lachat}  \&
  {Rasio}}{{Kremer} et~al.}{2019b}]{2019ApJ...881...75K}
{Kremer} K.,  {Lu} W.,  {Rodriguez} C.~L.,  {Lachat} M.,   {Rasio} F.~A.,
  2019b, \mn@doi [\apj] {10.3847/1538-4357/ab2e0c}, \href
  {https://ui.adsabs.harvard.edu/abs/2019ApJ...881...75K} {881, 75}

\bibitem[\protect\citeauthoryear{Lee, Ramirez-Ruiz  \& van~de Ven}{Lee
  et~al.}{2010}]{Lee:2010in}
Lee W.~H.,  Ramirez-Ruiz E.,   van~de Ven G.,  2010, \apj, 720, 953

\bibitem[\protect\citeauthoryear{{Leigh} et~al.,}{{Leigh}
  et~al.}{2018}]{2018MNRAS.474.5672L}
{Leigh} N.~W.~C.,  et~al., 2018, \mn@doi [\mnras] {10.1093/mnras/stx3134},
  \href {https://ui.adsabs.harvard.edu/abs/2018MNRAS.474.5672L} {474, 5672}

\bibitem[\protect\citeauthoryear{{Li} \& {Lai}}{{Li} \& {Lai}}{2022}]{Li22a}
{Li} R.,  {Lai} D.,  2022, \mn@doi [\mnras] {10.1093/mnras/stac2577}, \href
  {https://ui.adsabs.harvard.edu/abs/2022MNRAS.517.1602L} {517, 1602}

\bibitem[\protect\citeauthoryear{{Li} \& {Lai}}{{Li} \& {Lai}}{2023a}]{Li22c}
{Li} R.,  {Lai} D.,  2023a, \mn@doi [arXiv e-prints]
  {10.48550/arXiv.2303.12207}, \href
  {https://ui.adsabs.harvard.edu/abs/2023arXiv230312207L} {p. arXiv:2303.12207}

\bibitem[\protect\citeauthoryear{{Li} \& {Lai}}{{Li} \&
  {Lai}}{2023b}]{2023MNRAS.522.1881L}
{Li} R.,  {Lai} D.,  2023b, \mn@doi [\mnras] {10.1093/mnras/stad1117}, \href
  {https://ui.adsabs.harvard.edu/abs/2023MNRAS.522.1881L} {522, 1881}

\bibitem[\protect\citeauthoryear{{Li}, {Dempsey}, {Li}, {Lai}  \& {Li}}{{Li}
  et~al.}{2023}]{2023ApJ...944L..42L}
{Li} J.,  {Dempsey} A.~M.,  {Li} H.,  {Lai} D.,   {Li} S.,  2023, \mn@doi
  [\apjl] {10.3847/2041-8213/acb934}, \href
  {https://ui.adsabs.harvard.edu/abs/2023ApJ...944L..42L} {944, L42}

\bibitem[\protect\citeauthoryear{{Loeb}}{{Loeb}}{2016}]{Loeb:2016}
{Loeb} A.,  2016, \mn@doi [\apjl] {10.3847/2041-8205/819/2/L21}, \href
  {http://adsabs.harvard.edu/abs/2016ApJ...819L..21L} {819, L21}

\bibitem[\protect\citeauthoryear{{Lopez}, {Batta}, {Ramirez-Ruiz}, {Martinez}
  \& {Samsing}}{{Lopez} et~al.}{2019}]{2019ApJ...877...56L}
{Lopez} Martin J.,  {Batta} A.,  {Ramirez-Ruiz} E.,  {Martinez} I.,   {Samsing}
  J.,  2019, \mn@doi [\apj] {10.3847/1538-4357/ab1842}, \href
  {https://ui.adsabs.harvard.edu/abs/2019ApJ...877...56L} {877, 56}

\bibitem[\protect\citeauthoryear{{MacLeod} \& {Lin}}{{MacLeod} \&
  {Lin}}{2020}]{Macleod20}
{MacLeod} M.,  {Lin} D. N.~C.,  2020, \mn@doi [\apj]
  {10.3847/1538-4357/ab64db}, \href
  {https://ui.adsabs.harvard.edu/abs/2020ApJ...889...94M} {889, 94}

\bibitem[\protect\citeauthoryear{{McKernan}, {Ford}, {Lyra}  \&
  {Perets}}{{McKernan} et~al.}{2012}]{2012MNRAS.425..460M}
{McKernan} B.,  {Ford} K.~E.~S.,  {Lyra} W.,   {Perets} H.~B.,  2012, \mn@doi
  [\mnras] {10.1111/j.1365-2966.2012.21486.x}, \href
  {http://adsabs.harvard.edu/abs/2012MNRAS.425..460M} {425, 460}

\bibitem[\protect\citeauthoryear{{McKernan} et~al.,}{{McKernan}
  et~al.}{2018}]{2017arXiv170207818M}
{McKernan} B.,  et~al., 2018, \mn@doi [\apj] {10.3847/1538-4357/aadae5}, \href
  {https://ui.adsabs.harvard.edu/abs/2018ApJ...866...66M} {866, 66}

\bibitem[\protect\citeauthoryear{{Monaghan}}{{Monaghan}}{1976}]{1976MNRAS.177..583M}
{Monaghan} J.~J.,  1976, \mn@doi [\mnras] {10.1093/mnras/177.3.583}, \href
  {https://ui.adsabs.harvard.edu/abs/1976MNRAS.177..583M} {177, 583}

\bibitem[\protect\citeauthoryear{{Murguia-Berthier}, {MacLeod}, {Ramirez-Ruiz},
  {Antoni}  \& {Macias}}{{Murguia-Berthier} et~al.}{2017}]{2017ApJ...845..173M}
{Murguia-Berthier} A.,  {MacLeod} M.,  {Ramirez-Ruiz} E.,  {Antoni} A.,
  {Macias} P.,  2017, \mn@doi [\apj] {10.3847/1538-4357/aa8140}, \href
  {https://ui.adsabs.harvard.edu/abs/2017ApJ...845..173M} {845, 173}

\bibitem[\protect\citeauthoryear{{Nasim} et~al.,}{{Nasim}
  et~al.}{2023}]{Nasim22}
{Nasim} S.~S.,  et~al., 2023, \mn@doi [\mnras] {10.1093/mnras/stad1295}, \href
  {https://ui.adsabs.harvard.edu/abs/2023MNRAS.522.5393N} {522, 5393}

\bibitem[\protect\citeauthoryear{{Nitz} \& {Capano}}{{Nitz} \&
  {Capano}}{2021}]{2021ApJ...907L...9N}
{Nitz} A.~H.,  {Capano} C.~D.,  2021, \mn@doi [\apjl]
  {10.3847/2041-8213/abccc5}, \href
  {https://ui.adsabs.harvard.edu/abs/2021ApJ...907L...9N} {907, L9}

\bibitem[\protect\citeauthoryear{{O'Leary}, {Kocsis}  \& {Loeb}}{{O'Leary}
  et~al.}{2009}]{2009MNRAS.395.2127O}
{O'Leary} R.~M.,  {Kocsis} B.,   {Loeb} A.,  2009, \mn@doi [\mnras]
  {10.1111/j.1365-2966.2009.14653.x}, \href
  {http://adsabs.harvard.edu/abs/2009MNRAS.395.2127O} {395, 2127}

\bibitem[\protect\citeauthoryear{{O'Neill}, {D'Orazio}, {Samsing}  \&
  {Pessah}}{{O'Neill} et~al.}{2024}]{ONeill24}
{O'Neill} D.,  {D'Orazio} D.~J.,  {Samsing} J.,   {Pessah} M.~E.,  2024,
  \mn@doi [arXiv e-prints] {10.48550/arXiv.2401.16166}, \href
  {https://ui.adsabs.harvard.edu/abs/2024arXiv240116166O} {p. arXiv:2401.16166}

\bibitem[\protect\citeauthoryear{{Park}, {Kim}, {Lee}, {Bae}  \&
  {Belczynski}}{{Park} et~al.}{2017}]{2017MNRAS.469.4665P}
{Park} D.,  {Kim} C.,  {Lee} H.~M.,  {Bae} Y.-B.,   {Belczynski} K.,  2017,
  \mn@doi [\mnras] {10.1093/mnras/stx1015}, \href
  {http://adsabs.harvard.edu/abs/2017MNRAS.469.4665P} {469, 4665}

\bibitem[\protect\citeauthoryear{{Peters}}{{Peters}}{1964}]{Peters64}
{Peters} P.~C.,  1964, \mn@doi [Physical Review] {10.1103/PhysRev.136.B1224},
  \href {https://ui.adsabs.harvard.edu/abs/1964PhRv..136.1224P} {136, 1224}

\bibitem[\protect\citeauthoryear{{Piran} \& {Piran}}{{Piran} \&
  {Piran}}{2020}]{Piran2020ApJ}
{Piran} Z.,  {Piran} T.,  2020, \mn@doi [\apj] {10.3847/1538-4357/ab792a},
  \href {https://ui.adsabs.harvard.edu/abs/2020ApJ...892...64P} {892, 64}

\bibitem[\protect\citeauthoryear{{Portegies Zwart} \& {McMillan}}{{Portegies
  Zwart} \& {McMillan}}{2000}]{2000ApJ...528L..17P}
{Portegies Zwart} S.~F.,  {McMillan} S.~L.~W.,  2000, \mn@doi [\apjl]
  {10.1086/312422}, \href {http://adsabs.harvard.edu/abs/2000ApJ...528L..17P}
  {528, L17}

\bibitem[\protect\citeauthoryear{{Rodriguez} \& {Antonini}}{{Rodriguez} \&
  {Antonini}}{2018}]{2018ApJ...863....7R}
{Rodriguez} C.~L.,  {Antonini} F.,  2018, \mn@doi [\apj]
  {10.3847/1538-4357/aacea4}, \href
  {http://adsabs.harvard.edu/abs/2018ApJ...863....7R} {863, 7}

\bibitem[\protect\citeauthoryear{{Rodriguez}, {Morscher}, {Pattabiraman},
  {Chatterjee}, {Haster}  \& {Rasio}}{{Rodriguez}
  et~al.}{2015}]{2015PhRvL.115e1101R}
{Rodriguez} C.~L.,  {Morscher} M.,  {Pattabiraman} B.,  {Chatterjee} S.,
  {Haster} C.-J.,   {Rasio} F.~A.,  2015, \mn@doi [Physical Review Letters]
  {10.1103/PhysRevLett.115.051101}, \href
  {http://adsabs.harvard.edu/abs/2015PhRvL.115e1101R} {115, 051101}

\bibitem[\protect\citeauthoryear{{Rodriguez}, {Chatterjee}  \&
  {Rasio}}{{Rodriguez} et~al.}{2016a}]{2016PhRvD..93h4029R}
{Rodriguez} C.~L.,  {Chatterjee} S.,   {Rasio} F.~A.,  2016a, \mn@doi [\prd]
  {10.1103/PhysRevD.93.084029}, \href
  {http://adsabs.harvard.edu/abs/2016PhRvD..93h4029R} {93, 084029}

\bibitem[\protect\citeauthoryear{{Rodriguez}, {Haster}, {Chatterjee},
  {Kalogera}  \& {Rasio}}{{Rodriguez} et~al.}{2016b}]{2016ApJ...824L...8R}
{Rodriguez} C.~L.,  {Haster} C.-J.,  {Chatterjee} S.,  {Kalogera} V.,   {Rasio}
  F.~A.,  2016b, \mn@doi [\apjl] {10.3847/2041-8205/824/1/L8}, \href
  {http://adsabs.harvard.edu/abs/2016ApJ...824L...8R} {824, L8}

\bibitem[\protect\citeauthoryear{{Rodriguez}, {Zevin}, {Pankow}, {Kalogera}  \&
  {Rasio}}{{Rodriguez} et~al.}{2016c}]{2016ApJ...832L...2R}
{Rodriguez} C.~L.,  {Zevin} M.,  {Pankow} C.,  {Kalogera} V.,   {Rasio} F.~A.,
  2016c, \mn@doi [\apjl] {10.3847/2041-8205/832/1/L2}, \href
  {http://adsabs.harvard.edu/abs/2016ApJ...832L...2R} {832, L2}

\bibitem[\protect\citeauthoryear{{Rodriguez}, {Amaro-Seoane}, {Chatterjee},
  {Kremer}, {Rasio}, {Samsing}, {Ye}  \& {Zevin}}{{Rodriguez}
  et~al.}{2018}]{2018PhRvD..98l3005R}
{Rodriguez} C.~L.,  {Amaro-Seoane} P.,  {Chatterjee} S.,  {Kremer} K.,  {Rasio}
  F.~A.,  {Samsing} J.,  {Ye} C.~S.,   {Zevin} M.,  2018, \mn@doi [\prd]
  {10.1103/PhysRevD.98.123005}, \href
  {https://ui.adsabs.harvard.edu/abs/2018PhRvD..98l3005R} {98, 123005}

\bibitem[\protect\citeauthoryear{{Rom}, {Sari}  \& {Lai}}{{Rom}
  et~al.}{2023}]{2023arXiv231003801R}
{Rom} B.,  {Sari} R.,   {Lai} D.,  2023, \mn@doi [arXiv e-prints]
  {10.48550/arXiv.2310.03801}, \href
  {https://ui.adsabs.harvard.edu/abs/2023arXiv231003801R} {p. arXiv:2310.03801}

\bibitem[\protect\citeauthoryear{{Romero-Shaw}, {Lasky}, {Thrane}  \&
  {Calder{\'o}n Bustillo}}{{Romero-Shaw} et~al.}{2020}]{2020ApJ...903L...5R}
{Romero-Shaw} I.,  {Lasky} P.~D.,  {Thrane} E.,   {Calder{\'o}n Bustillo} J.,
  2020, \mn@doi [\apjl] {10.3847/2041-8213/abbe26}, \href
  {https://ui.adsabs.harvard.edu/abs/2020ApJ...903L...5R} {903, L5}

\bibitem[\protect\citeauthoryear{{Romero-Shaw}, {Lasky}  \&
  {Thrane}}{{Romero-Shaw} et~al.}{2022}]{2022ApJ...940..171R}
{Romero-Shaw} I.,  {Lasky} P.~D.,   {Thrane} E.,  2022, \mn@doi [\apj]
  {10.3847/1538-4357/ac9798}, \href
  {https://ui.adsabs.harvard.edu/abs/2022ApJ...940..171R} {940, 171}

\bibitem[\protect\citeauthoryear{{Rowan}, {Boekholt}, {Kocsis}  \&
  {Haiman}}{{Rowan} et~al.}{2023}]{Rowan23b}
{Rowan} C.,  {Boekholt} T.,  {Kocsis} B.,   {Haiman} Z.,  2023, \mn@doi
  [\mnras] {10.1093/mnras/stad1926}, \href
  {https://ui.adsabs.harvard.edu/abs/2023MNRAS.524.2770R} {524, 2770}

\bibitem[\protect\citeauthoryear{{Rowan}, {Whitehead}, {Boekholt}, {Kocsis}  \&
  {Haiman}}{{Rowan} et~al.}{2024}]{Rowan24a}
{Rowan} C.,  {Whitehead} H.,  {Boekholt} T.,  {Kocsis} B.,   {Haiman} Z.,
  2024, \mn@doi [\mnras] {10.1093/mnras/stad3641}, \href
  {https://ui.adsabs.harvard.edu/abs/2024MNRAS.52710448R} {527, 10448}

\bibitem[\protect\citeauthoryear{{Rozner} \& {Perets}}{{Rozner} \&
  {Perets}}{2022}]{2022ApJ...931..149R}
{Rozner} M.,  {Perets} H.~B.,  2022, \mn@doi [\apj] {10.3847/1538-4357/ac6d55},
  \href {https://ui.adsabs.harvard.edu/abs/2022ApJ...931..149R} {931, 149}

\bibitem[\protect\citeauthoryear{{Rozner}, {Generozov}  \& {Perets}}{{Rozner}
  et~al.}{2023}]{2023MNRAS.521..866R}
{Rozner} M.,  {Generozov} A.,   {Perets} H.~B.,  2023, \mn@doi [\mnras]
  {10.1093/mnras/stad603}, \href
  {https://ui.adsabs.harvard.edu/abs/2023MNRAS.521..866R} {521, 866}

\bibitem[\protect\citeauthoryear{{Samsing}}{{Samsing}}{2018}]{Samsing18}
{Samsing} J.,  2018, \mn@doi [\prd] {10.1103/PhysRevD.97.103014}, \href
  {https://ui.adsabs.harvard.edu/abs/2018PhRvD..97j3014S} {97, 103014}

\bibitem[\protect\citeauthoryear{{Samsing} \& {D'Orazio}}{{Samsing} \&
  {D'Orazio}}{2018}]{2018MNRAS.tmp.2223S}
{Samsing} J.,  {D'Orazio} D.~J.,  2018, \mn@doi [\mnras]
  {10.1093/mnras/sty2334}, \href
  {http://adsabs.harvard.edu/abs/2018MNRAS.tmp.2223S} {}

\bibitem[\protect\citeauthoryear{{Samsing} \& {Ilan}}{{Samsing} \&
  {Ilan}}{2018}]{Samsing18a}
{Samsing} J.,  {Ilan} T.,  2018, \mn@doi [\mnras] {10.1093/mnras/sty197}, \href
  {https://ui.adsabs.harvard.edu/abs/2018MNRAS.476.1548S} {476, 1548}

\bibitem[\protect\citeauthoryear{{Samsing} \& {Ramirez-Ruiz}}{{Samsing} \&
  {Ramirez-Ruiz}}{2017}]{2017ApJ...840L..14S}
{Samsing} J.,  {Ramirez-Ruiz} E.,  2017, \mn@doi [\apjl]
  {10.3847/2041-8213/aa6f0b}, \href
  {http://adsabs.harvard.edu/abs/2017ApJ...840L..14S} {840, L14}

\bibitem[\protect\citeauthoryear{{Samsing}, {MacLeod}  \&
  {Ramirez-Ruiz}}{{Samsing} et~al.}{2014}]{Samsing14}
{Samsing} J.,  {MacLeod} M.,   {Ramirez-Ruiz} E.,  2014, \mn@doi [\apj]
  {10.1088/0004-637X/784/1/71}, \href
  {https://ui.adsabs.harvard.edu/abs/2014ApJ...784...71S} {784, 71}

\bibitem[\protect\citeauthoryear{{Samsing}, {MacLeod}  \&
  {Ramirez-Ruiz}}{{Samsing} et~al.}{2017}]{Samsing17}
{Samsing} J.,  {MacLeod} M.,   {Ramirez-Ruiz} E.,  2017, \mn@doi [\apj]
  {10.3847/1538-4357/aa7e32}, \href
  {https://ui.adsabs.harvard.edu/abs/2017ApJ...846...36S} {846, 36}

\bibitem[\protect\citeauthoryear{{Samsing}, {MacLeod}  \&
  {Ramirez-Ruiz}}{{Samsing} et~al.}{2018a}]{Samsing2018}
{Samsing} J.,  {MacLeod} M.,   {Ramirez-Ruiz} E.,  2018a, \mn@doi [\apj]
  {10.3847/1538-4357/aaa715}, \href
  {https://ui.adsabs.harvard.edu/abs/2018ApJ...853..140S} {853, 140}

\bibitem[\protect\citeauthoryear{{Samsing}, {Askar}  \& {Giersz}}{{Samsing}
  et~al.}{2018b}]{2018ApJ...855..124S}
{Samsing} J.,  {Askar} A.,   {Giersz} M.,  2018b, \mn@doi [\apj]
  {10.3847/1538-4357/aaab52}, \href
  {http://adsabs.harvard.edu/abs/2018ApJ...855..124S} {855, 124}

\bibitem[\protect\citeauthoryear{{Samsing}, {Venumadhav}, {Dai}, {Martinez},
  {Batta}, {Lopez}, {Ramirez-Ruiz}  \& {Kremer}}{{Samsing}
  et~al.}{2019a}]{2019PhRvD.100d3009S}
{Samsing} J.,  {Venumadhav} T.,  {Dai} L.,  {Martinez} I.,  {Batta} A.,
  {Lopez} M.,  {Ramirez-Ruiz} E.,   {Kremer} K.,  2019a, \mn@doi [\prd]
  {10.1103/PhysRevD.100.043009}, \href
  {https://ui.adsabs.harvard.edu/abs/2019PhRvD.100d3009S} {100, 043009}

\bibitem[\protect\citeauthoryear{{Samsing}, {Hamers}  \& {Tyles}}{{Samsing}
  et~al.}{2019b}]{2019PhRvD.100d3010S}
{Samsing} J.,  {Hamers} A.~S.,   {Tyles} J.~G.,  2019b, \mn@doi [\prd]
  {10.1103/PhysRevD.100.043010}, \href
  {https://ui.adsabs.harvard.edu/abs/2019PhRvD.100d3010S} {100, 043010}

\bibitem[\protect\citeauthoryear{{Samsing}, {D'Orazio}, {Kremer}, {Rodriguez}
  \& {Askar}}{{Samsing} et~al.}{2020}]{2019arXiv190711231S}
{Samsing} J.,  {D'Orazio} D.~J.,  {Kremer} K.,  {Rodriguez} C.~L.,   {Askar}
  A.,  2020, \mn@doi [\prd] {10.1103/PhysRevD.101.123010}, \href
  {https://ui.adsabs.harvard.edu/abs/2020PhRvD.101l3010S} {101, 123010}

\bibitem[\protect\citeauthoryear{{Samsing} et~al.,}{{Samsing}
  et~al.}{2022}]{Samsing22}
{Samsing} J.,  et~al., 2022, \mn@doi [\nat] {10.1038/s41586-021-04333-1}, \href
  {https://ui.adsabs.harvard.edu/abs/2022Natur.603..237S} {603, 237}

\bibitem[\protect\citeauthoryear{{Samsing}, {Hendriks}, {Zwick}, {D'Orazio}  \&
  {Liu}}{{Samsing} et~al.}{2024}]{2024arXiv240305625S}
{Samsing} J.,  {Hendriks} K.,  {Zwick} L.,  {D'Orazio} D.~J.,   {Liu} B.,
  2024, \mn@doi [arXiv e-prints] {10.48550/arXiv.2403.05625}, \href
  {https://ui.adsabs.harvard.edu/abs/2024arXiv240305625S} {p. arXiv:2403.05625}

\bibitem[\protect\citeauthoryear{{Sasaki}, {Suyama}, {Tanaka}  \&
  {Yokoyama}}{{Sasaki} et~al.}{2016}]{2016PhRvL.117f1101S}
{Sasaki} M.,  {Suyama} T.,  {Tanaka} T.,   {Yokoyama} S.,  2016, \mn@doi
  [Physical Review Letters] {10.1103/PhysRevLett.117.061101}, \href
  {http://adsabs.harvard.edu/abs/2016PhRvL.117f1101S} {117, 061101}

\bibitem[\protect\citeauthoryear{{Schr{\o}der}, {Batta}  \&
  {Ramirez-Ruiz}}{{Schr{\o}der} et~al.}{2018}]{2018ApJ...862L...3S}
{Schr{\o}der} S.~L.,  {Batta} A.,   {Ramirez-Ruiz} E.,  2018, \mn@doi [\apjl]
  {10.3847/2041-8213/aacf8d}, \href
  {https://ui.adsabs.harvard.edu/abs/2018ApJ...862L...3S} {862, L3}

\bibitem[\protect\citeauthoryear{{Secunda}, {Bellovary}, {Mac Low}, {Ford},
  {McKernan}, {Leigh}, {Lyra}  \& {S{\'a}ndor}}{{Secunda}
  et~al.}{2019}]{2019ApJ...878...85S}
{Secunda} A.,  {Bellovary} J.,  {Mac Low} M.-M.,  {Ford} K.~E.~S.,  {McKernan}
  B.,  {Leigh} N. W.~C.,  {Lyra} W.,   {S{\'a}ndor} Z.,  2019, \mn@doi [\apj]
  {10.3847/1538-4357/ab20ca}, \href
  {https://ui.adsabs.harvard.edu/abs/2019ApJ...878...85S} {878, 85}

\bibitem[\protect\citeauthoryear{{Secunda} et~al.,}{{Secunda}
  et~al.}{2020}]{2020ApJ...903..133S}
{Secunda} A.,  et~al., 2020, \mn@doi [\apj] {10.3847/1538-4357/abbc1d}, \href
  {https://ui.adsabs.harvard.edu/abs/2020ApJ...903..133S} {903, 133}

\bibitem[\protect\citeauthoryear{{Secunda}, {Hernandez}, {Goodman}, {Leigh},
  {McKernan}, {Ford}  \& {Adorno}}{{Secunda}
  et~al.}{2021}]{2021ApJ...908L..27S}
{Secunda} A.,  {Hernandez} B.,  {Goodman} J.,  {Leigh} N. W.~C.,  {McKernan}
  B.,  {Ford} K.~E.~S.,   {Adorno} J.~I.,  2021, \mn@doi [\apjl]
  {10.3847/2041-8213/abe11d}, \href
  {https://ui.adsabs.harvard.edu/abs/2021ApJ...908L..27S} {908, L27}

\bibitem[\protect\citeauthoryear{{Shakura} \& {Sunyaev}}{{Shakura} \&
  {Sunyaev}}{1973}]{Shakura73}
{Shakura} N.~I.,  {Sunyaev} R.~A.,  1973, \aap, \href
  {https://ui.adsabs.harvard.edu/abs/1973A&A....24..337S} {24, 337}

\bibitem[\protect\citeauthoryear{{Silsbee} \& {Tremaine}}{{Silsbee} \&
  {Tremaine}}{2017}]{2017ApJ...836...39S}
{Silsbee} K.,  {Tremaine} S.,  2017, \mn@doi [\apj] {10.3847/1538-4357/aa5729},
  \href {http://adsabs.harvard.edu/abs/2017ApJ...836...39S} {836, 39}

\bibitem[\protect\citeauthoryear{{Sirko} \& {Goodman}}{{Sirko} \&
  {Goodman}}{2003}]{SG}
{Sirko} E.,  {Goodman} J.,  2003, \mn@doi [\mnras]
  {10.1046/j.1365-8711.2003.06431.x}, \href
  {https://ui.adsabs.harvard.edu/abs/2003MNRAS.341..501S} {341, 501}

\bibitem[\protect\citeauthoryear{{Stephan}, {Naoz}, {Ghez}, {Witzel},
  {Sitarski}, {Do}  \& {Kocsis}}{{Stephan} et~al.}{2016}]{2016MNRAS.460.3494S}
{Stephan} A.~P.,  {Naoz} S.,  {Ghez} A.~M.,  {Witzel} G.,  {Sitarski} B.~N.,
  {Do} T.,   {Kocsis} B.,  2016, \mn@doi [\mnras] {10.1093/mnras/stw1220},
  \href {http://adsabs.harvard.edu/abs/2016MNRAS.460.3494S} {460, 3494}

\bibitem[\protect\citeauthoryear{{Stone} \& {Leigh}}{{Stone} \&
  {Leigh}}{2019}]{2019Natur.576..406S}
{Stone} N.~C.,  {Leigh} N. W.~C.,  2019, \mn@doi [\nat]
  {10.1038/s41586-019-1833-8}, \href
  {https://ui.adsabs.harvard.edu/abs/2019Natur.576..406S} {576, 406}

\bibitem[\protect\citeauthoryear{{Stone}, {Metzger}  \& {Haiman}}{{Stone}
  et~al.}{2017}]{2017MNRAS.464..946S}
{Stone} N.~C.,  {Metzger} B.~D.,   {Haiman} Z.,  2017, \mn@doi [\mnras]
  {10.1093/mnras/stw2260}, \href
  {http://adsabs.harvard.edu/abs/2017MNRAS.464..946S} {464, 946}

\bibitem[\protect\citeauthoryear{{Tagawa}, {Haiman}  \& {Kocsis}}{{Tagawa}
  et~al.}{2019}]{2019arXiv191208218T}
{Tagawa} H.,  {Haiman} Z.,   {Kocsis} B.,  2019, arXiv e-prints, \href
  {https://ui.adsabs.harvard.edu/abs/2019arXiv191208218T} {p. arXiv:1912.08218}

\bibitem[\protect\citeauthoryear{{Tagawa}, {Haiman}  \& {Kocsis}}{{Tagawa}
  et~al.}{2020}]{Tagawa20}
{Tagawa} H.,  {Haiman} Z.,   {Kocsis} B.,  2020, \mn@doi [\apj]
  {10.3847/1538-4357/ab9b8c}, \href
  {https://ui.adsabs.harvard.edu/abs/2020ApJ...898...25T} {898, 25}

\bibitem[\protect\citeauthoryear{{Tagawa}, {Kocsis}, {Haiman}, {Bartos},
  {Omukai}  \& {Samsing}}{{Tagawa} et~al.}{2021}]{Tagawa21}
{Tagawa} H.,  {Kocsis} B.,  {Haiman} Z.,  {Bartos} I.,  {Omukai} K.,
  {Samsing} J.,  2021, \mn@doi [\apj] {10.3847/1538-4357/abd555}, \href
  {https://ui.adsabs.harvard.edu/abs/2021ApJ...908..194T} {908, 194}

\bibitem[\protect\citeauthoryear{{Tanikawa}}{{Tanikawa}}{2013}]{2013MNRAS.435.1358T}
{Tanikawa} A.,  2013, \mn@doi [\mnras] {10.1093/mnras/stt1380}, \href
  {http://adsabs.harvard.edu/abs/2013MNRAS.435.1358T} {435, 1358}

\bibitem[\protect\citeauthoryear{{Thompson}, {Quataert}  \&
  {Murray}}{{Thompson} et~al.}{2005}]{TQM}
{Thompson} T.~A.,  {Quataert} E.,   {Murray} N.,  2005, \mn@doi [\apj]
  {10.1086/431923}, \href
  {https://ui.adsabs.harvard.edu/abs/2005ApJ...630..167T} {630, 167}

\bibitem[\protect\citeauthoryear{{Trani}, {Spera}, {Leigh}  \& {Fujii}}{{Trani}
  et~al.}{2019}]{2019ApJ...885..135T}
{Trani} A.~A.,  {Spera} M.,  {Leigh} N. W.~C.,   {Fujii} M.~S.,  2019, \mn@doi
  [\apj] {10.3847/1538-4357/ab480a}, \href
  {https://ui.adsabs.harvard.edu/abs/2019ApJ...885..135T} {885, 135}

\bibitem[\protect\citeauthoryear{{Trani}, {Quaini}  \& {Colpi}}{{Trani}
  et~al.}{2023}]{Trani23}
{Trani} A.~A.,  {Quaini} S.,   {Colpi} M.,  2023, \mn@doi [arXiv e-prints]
  {10.48550/arXiv.2312.13281}, \href
  {https://ui.adsabs.harvard.edu/abs/2023arXiv231213281T} {p. arXiv:2312.13281}

\bibitem[\protect\citeauthoryear{{Valli} et~al.,}{{Valli}
  et~al.}{2024}]{2024arXiv240117355V}
{Valli} R.,  et~al., 2024, \mn@doi [arXiv e-prints]
  {10.48550/arXiv.2401.17355}, \href
  {https://ui.adsabs.harvard.edu/abs/2024arXiv240117355V} {p. arXiv:2401.17355}

\bibitem[\protect\citeauthoryear{{Valtonen} \& {Karttunen}}{{Valtonen} \&
  {Karttunen}}{2006}]{2006tbp..book.....V}
{Valtonen} M.,  {Karttunen} H.,  2006, {The Three-Body Problem}

\bibitem[\protect\citeauthoryear{{VanLandingham}, {Miller}, {Hamilton}  \&
  {Richardson}}{{VanLandingham} et~al.}{2016}]{2016ApJ...828...77V}
{VanLandingham} J.~H.,  {Miller} M.~C.,  {Hamilton} D.~P.,   {Richardson}
  D.~C.,  2016, \mn@doi [\apj] {10.3847/0004-637X/828/2/77}, \href
  {http://adsabs.harvard.edu/abs/2016ApJ...828...77V} {828, 77}

\bibitem[\protect\citeauthoryear{{Venumadhav}, {Zackay}, {Roulet}, {Dai}  \&
  {Zaldarriaga}}{{Venumadhav} et~al.}{2020a}]{2019arXiv190210331Z}
{Venumadhav} T.,  {Zackay} B.,  {Roulet} J.,  {Dai} L.,   {Zaldarriaga} M.,
  2020a, \mn@doi [\prd] {10.1103/PhysRevD.101.083030}, \href
  {https://ui.adsabs.harvard.edu/abs/2020PhRvD.101h3030V} {101, 083030}

\bibitem[\protect\citeauthoryear{{Venumadhav}, {Zackay}, {Roulet}, {Dai}  \&
  {Zaldarriaga}}{{Venumadhav} et~al.}{2020b}]{2019arXiv190407214V}
{Venumadhav} T.,  {Zackay} B.,  {Roulet} J.,  {Dai} L.,   {Zaldarriaga} M.,
  2020b, \mn@doi [\prd] {10.1103/PhysRevD.101.083030}, \href
  {https://ui.adsabs.harvard.edu/abs/2020PhRvD.101h3030V} {101, 083030}

\bibitem[\protect\citeauthoryear{{Vynatheya} \& {Hamers}}{{Vynatheya} \&
  {Hamers}}{2022}]{Vyntantheya22}
{Vynatheya} P.,  {Hamers} A.~S.,  2022, \mn@doi [\apj]
  {10.3847/1538-4357/ac4892}, \href
  {https://ui.adsabs.harvard.edu/abs/2022ApJ...926..195V} {926, 195}

\bibitem[\protect\citeauthoryear{{Wang}, {Zhu}  \& {Lin}}{{Wang}
  et~al.}{2024}]{2024MNRAS.528.4958W}
{Wang} Y.,  {Zhu} Z.,   {Lin} D. N.~C.,  2024, \mn@doi [\mnras]
  {10.1093/mnras/stae321}, \href
  {https://ui.adsabs.harvard.edu/abs/2024MNRAS.528.4958W} {528, 4958}

\bibitem[\protect\citeauthoryear{{Whitehead}, {Rowan}, {Boekholt}  \&
  {Kocsis}}{{Whitehead} et~al.}{2024}]{2024MNRAS.533.1766W}
{Whitehead} H.,  {Rowan} C.,  {Boekholt} T.,   {Kocsis} B.,  2024, \mn@doi
  [\mnras] {10.1093/mnras/stae1866}, \href
  {https://ui.adsabs.harvard.edu/abs/2024MNRAS.533.1766W} {533, 1766}

\bibitem[\protect\citeauthoryear{{Woosley}}{{Woosley}}{2016}]{Woosley:2016}
{Woosley} S.~E.,  2016, \mn@doi [\apjl] {10.3847/2041-8205/824/1/L10}, \href
  {http://adsabs.harvard.edu/abs/2016ApJ...824L..10W} {824, L10}

\bibitem[\protect\citeauthoryear{{Zaldarriaga}, {Kushnir}  \&
  {Kollmeier}}{{Zaldarriaga} et~al.}{2018}]{Zaldarriaga2018MNRAS}
{Zaldarriaga} M.,  {Kushnir} D.,   {Kollmeier} J.~A.,  2018, \mn@doi [\mnras]
  {10.1093/mnras/stx2577}, \href
  {https://ui.adsabs.harvard.edu/abs/2018MNRAS.473.4174Z} {473, 4174}

\bibitem[\protect\citeauthoryear{{Zevin}, {Pankow}, {Rodriguez}, {Sampson},
  {Chase}, {Kalogera}  \& {Rasio}}{{Zevin} et~al.}{2017}]{2017ApJ...846...82Z}
{Zevin} M.,  {Pankow} C.,  {Rodriguez} C.~L.,  {Sampson} L.,  {Chase} E.,
  {Kalogera} V.,   {Rasio} F.~A.,  2017, \mn@doi [\apj]
  {10.3847/1538-4357/aa8408}, \href
  {https://ui.adsabs.harvard.edu/#abs/2017ApJ...846...82Z} {846, 82}

\bibitem[\protect\citeauthoryear{{Zevin}, {Samsing}, {Rodriguez}, {Haster}  \&
  {Ramirez-Ruiz}}{{Zevin} et~al.}{2019}]{2019ApJ...871...91Z}
{Zevin} M.,  {Samsing} J.,  {Rodriguez} C.,  {Haster} C.-J.,   {Ramirez-Ruiz}
  E.,  2019, \mn@doi [\apj] {10.3847/1538-4357/aaf6ec}, \href
  {https://ui.adsabs.harvard.edu/abs/2019ApJ...871...91Z} {871, 91}

\makeatother
\end{thebibliography}

\bsp
\label{lastpage}

\end{document}